\documentclass[12pt,twoside]{article}

\usepackage[mathscr]{eucal}
\usepackage{amsmath,amsfonts,amssymb,amsthm}
\usepackage{times}
\usepackage{srcltx}
\usepackage{hyperref}

\advance\textheight\topskip
\numberwithin{equation}{section} \makeatletter
\@addtoreset{equation}{section}

\renewcommand{\tilde}{\widetilde}
\renewcommand{\hat}{\widehat}

\newcommand{\bref}[1]{\textbf{\ref{#1}}}

\renewcommand{\mod}{\,\rm mod \,}
\newcommand{\p}[1]{|#1|}
\newcommand{\gh}[1]{\mathrm{gh}(#1)}

\newcommand{\dd}{\partial}
\renewcommand{\d}{\partial}

\renewcommand{\geq}{\,{\geqslant}\,}
\renewcommand{\leq}{\,{\leqslant}\,}

\newcommand{\inner}[2]{\langle #1{,}\,#2\rangle}
\newcommand{\binner}[2]{%
  {\langle}\kern-4.15pt{\langle}#1{,}\,#2{\rangle}\kern-4.15pt{\rangle}}
\newcommand{\commut}[2]{[#1{,}\,#2]}
\newcommand{\qcommut}[2]{[#1{,}\,#2]_*}
\newcommand{\pb}[2]{\left\{{}#1{},{}#2{}\right\}}
\newcommand{\ab}[2]{\big(#1,#2\big)}

\newcommand{\half}{\mathchoice{%
    \ffrac{1}{2}}{\frac{1}{2}}{\frac{1}{2}}{\frac{1}{2}}}

\newcommand{\ffrac}[2]{\raisebox{.5pt}%
  {\footnotesize$\displaystyle\frac{#1}{#2}$}\kern1pt}

\newcommand{\tot}{\mathrm{T}}
\newcommand{\red}{\mathrm{red}}

\newcommand{\derham}{\boldsymbol{d}}

\newcommand{\dl}[1]{\mathchoice{\ffrac{\dd}{\dd #1}}{\frac{\dd}{\dd
      #1}}{\ffrac{\dd}{\dd #1}}{\ffrac{\dd}{\dd #1}}}

\newcommand{\ddl}[2]{\ffrac{\dd #1}{\dd #2}}

\newcommand{\vdl}[1]{\ffrac{{\delta}}{\delta #1}}

\newcommand{\vddr}[2]{\ffrac{\delta^R #1}{\delta #2}}
\newcommand{\vddl}[2]{{\ffrac{\delta #1}{\delta #2}}}

\def\const{\mathop\mathrm{const}\nolimits}

\newcommand{\manifold}[1]{\mathscr{#1}}
\newcommand{\manX}{\manifold{X}}

\newcommand{\manM}{\manifold{M}}
\newcommand{\manN}{\manifold{N}}
\newcommand{\manZ}{\manifold{Z}}

\newcommand{\Liealg}{\mathfrak} 
\newcommand{\algg}{\Liealg{g}}

\newcommand{\fR}{\mathbb{R}}

\def\cA{\mathcal{A}}

\def\cC{\mathcal{C}}

\def\cF{\mathcal{F}}

\def\cP{\mathcal{P}}

\newcommand{\Vol}{\mathcal{V}}

\def\sd{S^\dagger}

\def\BGST{Barnich:2004cr}
\def\BGadS{Barnich:2006pc}

\def\Goff{Grigoriev:2006tt}

\def\AGT{Alkalaev:2008gi}
\def\GD{Grigoriev:1999qz}
\def\BGnlp{Barnich:2010sw}
\def\BG-Poincare{Barnich:2009jy}
\def\Fedosov-book{Fedosov:1996fu}

\begin{document}

 \begin{flushright}\small
 FIAN-TD-2012-08
 \end{flushright}

\vspace{1cm}
\begin{centering}

  \vspace{1cm}

\textbf{\Large{Parent formulations, frame-like Lagrangians, and generalized auxiliary fields}}

  \vspace{1cm}

  {\large Maxim Grigoriev} 

\vspace{.7cm}

\begin{minipage}{.9\textwidth} \it \begin{center}
   Tamm Theory Department, Lebedev Physics
   Institute,\\ Leninsky prospect 53,  119991 Moscow, Russia\end{center}
\end{minipage}

\end{centering}

\vspace{2cm}

\begin{center}
  \begin{minipage}{.88\textwidth}
    \textsc{Abstract}. 
    We elaborate on the recently proposed Lagrangian parent formulation. In particular, we identify a natural choice of the allowed field configurations ensuring the equivalence of the parent
and the starting point Lagrangians. We also analyze the structure of the generalized auxiliary fields employed in the parent formulation and establish the relationship between the parent Lagrangian and the recently proposed Lagrange structure for the unfolded dynamics. As an illustration of the parent formalism a systematic derivation of the frame-like Lagrangian for totally symmetric fields starting from the Fronsdal one is given.
We also present a concise and manifestly $sp(2)$-symmetric form of the off-shell constraints and gauge symmetries for AdS higher spin fields at the nonlinear level. 
\end{minipage}
\end{center}
\vspace{1cm}
\tableofcontents
\section{Introduction}

Equations of motion of a given field theory can be represented as a free differential algebra (FDA) through the inclusion of (usually infinite number of) auxiliary and St\"ueckelberg fields. Such form is referred to as unfolded and has been proved especially useful in the context of higher spin gauge theories. In particular, the interacting theory of higher spin fields on AdS background has been constructed~\cite{Vasiliev:1988sa,Vasiliev:1992av,Vasiliev:2003ev} in this framework.
The unfolded approach~\cite{Vasiliev:1988xc,Vasiliev:2005zu} is also a powerful tool in studying gauge field theories invariant under one or another space-time symmetry algebras~\cite{Vasiliev:2001zy,Shaynkman:2004vu}.

As far as general gauge theories are concerned the well-established framework is provided by the Batalin-Vilkovisky (BV) formalism~\cite{Batalin:1981jr,Batalin:1983wj}. In the context of local gauge theories the formalism operates in terms of the appropriate jet-bundles~\cite{DuboisViolette:1985jb,Barnich:1995db,Piguet:1995er} (see also~\cite{Andersonbook,Dickey:1991xa,Olver:1993,vinogradov:2001,Krasil'shchik:2010ij}
for the general mathematical introduction to jet-bundles and
partial differential equations).  In this approach basic objects such as the BRST differential, horizontal differential, conserved currents,  etc. become geometric objects on the jet-bundle. The BV approach
provides an efficient homological technique for studying renormalziation, consistent deformations, global symmetries etc. ~\cite{Piguet:1995er,Barnich:1993vg,Barnich:2000zw}.

Although there are definite similarities between the jet-space BV and the unfolded approach the detailed relationship
is not so obvious. At the level of equations of motion the relation between the BV
formalism and the unfolded approach was established in~\cite{\BGST}
(see also~\cite{\BGadS,Barnich:2005ru}) for linear systems and in
\cite{\BGnlp} in the general case by constructing the so-called parent
formulation such that both the BV and the unfolded formulation can be
arrived through different reductions of the parent one. In particular,
this construction gives a systematic way to derive the unfolded form of a given 
gauge theory.

The Lagrangian counterpart of the parent formulation is also known by now.
It has the form of an AKSZ sigma model~\cite{Alexandrov:1997kv} (see also~\cite{Cattaneo:1999fm,%
  Grigoriev:1999qz,Batalin:2001fc, Park:2000au,Roytenberg:2002nu,Kazinski:2005eb,
  Barnich:2009jy} for more recent developments and \cite{Baulieu:1995bq, Dayi:2003xx,Kotov:2010wr} for related constructions) whose target space is an appropriate
graded cotangent bundle over the supermanifold equipped with the nilpotent
differential $\tilde\gamma$ and the $\tilde\gamma$-invariant Lagrange potential.
For a given theory the supermanifold is the jet-space associated to fields, ghosts, ghosts-for-ghosts etc. but the antifields.  The genuine BV antifields are present among AKSZ sigma model fields whose space carries the BV antibracket induced by the target space (odd) symplectic structure.

The parent approach can be used as a tool to study hidden geometry of the theory
and to derive formulations which manifest one or another structure or symmetry.
For instance, starting with a parent formulation of a given theory and eliminating a certain subset of generalized auxiliary fields one identifies generalized connections and curvatures of~\cite{Brandt:1996mh,Brandt:1997iu}
and reformulates the theory in their terms. Note that these structures were also independently identified
within the unfolded approach from a slightly different perspective. Further reduction typically results in a frame-like formulation. In particular, it was demonstrated in~\cite{Grigoriev:2010ic,\BGnlp} that starting from
the usual metric-like formulation of gravity one systematically derives its Cartan--Weyl
formulation in terms of the frame field and Lorentz connection along with the familiar frame-like Lagrangian. Similar analysis reproduces the well-known first-order formulations for scalar and Yang-Mills fields.

Instead of using parent formulation to derive unfolded or other forms of an
already given theory one can look for new systems and analyze them using the
parent-like formulation as a starting point. In this way this becomes a powerful
framework unifying the ideas and methods of both jet-space BV and the unfolded
approach.  This strategy has proved fruitful in studying general gauge fields on
constant curvature spaces~\cite{\AGT,Bekaert:2009fg,Alkalaev:2009vm}. In this
context the equations of motion version of the parent approach has been used
from the very beginning to derive the equations of motion and gauge symmetries
in the concise and tractable form. One or another version of the parent
Lagrangian approach seem inevitable at the quantum level because it contains all
the relevant structures of the BV formalism (and, in fact, its Hamiltonian
analog known as Batalin--Fradkin--Vilkovisky (BFV)
quantization~\cite{Fradkin:1975cq,Batalin:1977pb,Fradkin:1978xi}) and hence
provides a natural setup for quantization.

The paper is organized as follows. Relevant information on jet-spaces,
generalized auxiliary fields and AKSZ sigma models is collected in
Section~\bref{sec:prelim}. Then in Section~\bref{sec:parent-main} we recall the
construction of the Lagrangian parent formulation for a given gauge theory. In
contrast to~\cite{Grigoriev:2010ic} we immediately start with the parametrized
version which makes the exposition more compact and geometrical. We then propose
the precise specification for the space of allowed field configurations which
guarantees the equivalence without the artificial truncation originally employed
in~\cite{Grigoriev:2010ic}. The relation to the conventional unfolded approach
is discussed in Section~\bref{sec:structure}. We then show that the recently
proposed canonical Lagrange structure for the unfolded
equations~\cite{Kaparulin:2011zz} can be systematically obtained from the
respective parent Lagrangian. As a byproduct this gives a systematic way to
explicitly construct canonical Lagrange structures for the unfolded equations. 
We complete the discussion by studying various subtleties emerging if the number of generalized auxiliary fields is infinite.

In Section~\bref{sec:HS} we demonstrate how the formalism works using free
totally symmetric gauge fields~\cite{Fronsdal:1978rb} as an example. Starting
with the Fronsdal Lagrangian we construct a version of the parent formulation
where the Lagrangian potential is relatively simple thanks to the parent version
of the gauge where the field is traceless and the gauge parameter is
transverse~\cite{Alvarez:2006uu,Blas:2007pp,Skvortsov:2007kz}. A remarkable
feature is that in contrast to the usual treatment this gauge is purely
algebraic in the parent setting. We then explicitly show that eliminating
further auxiliary fields results in the familiar frame like Lagrangian
of~\cite{Vasiliev:1980as}. We hope that understanding the structures underlying the
Lagrangian formulation of higher spin fields will be helpful at the nonlinear level as well
where despite of interesting developments~\cite{Boulanger:2011dd,Colombo:2010fu}
the problem of proper Lagrangian formulation of Vasiliev system~\cite{Vasiliev:2003ev} is still open.

As a next illustration of the parent approach in Section~\bref{sec:off} 
we present a concise form of the nonlinear off-shell constraints and gauge symmetries for AdS higher spins where
the familiar $sp(2)$-symmetry and local AdS symmetry are manifestly realized. Using
the parent formalism is essential in this context because the target space involves
negative degree coordinates and hence the equations of motion are not of the FDA
form. This formulation is deeply related to a quantized particle model for which the HS fields form a background~\cite{Segal:2002gd,\Goff} and can be seen as a generalization of
the remarkable flat-space system~\cite{Vasiliev:2005zu} or its AdS space version~\cite{\Goff}.

\section{Preliminaries}
\label{sec:prelim}

\subsection{Generalized auxiliary fields}
\label{sec:af}
Let us briefly recall the jet space language for local gauge field theories defined in the BRST theory terms.
More detailed account of the jet-bundle approach can be found in e.g.~\cite{Barnich:2000zw} (see also~\cite{Andersonbook,Dickey:1991xa,Olver:1993,vinogradov:2001,Krasil'shchik:2010ij} for the general mathematical introduction to jet-bundles and
partial differential equations). Let $\psi^A$ denote fields of the theory including ghosts, antifields etc. and $z^a$ denote space-time coordinates $a=0,1,\ldots,n-1$. Fields carry an integer ghost degree denoted by $\gh{}$.
Physical fields are found at vanishing ghost degree while fields of nonzero degree are interpreted as ghosts
and antifields associated to gauge symmetries, equations of motion, and their reducibilty relations.
The Grassmann parity of $\psi^A$ is denoted by $\p{\psi^A}$. For all the objects we need in the paper $\p{F}$ is always $\gh{F} \mod 2$ as for simplicity we do not explicitly consider systems with physical fermions.

The jet space associated to the space of fields is a supermanifold with coordinates $z^a,\xi^a\equiv dz^a, \psi^A$
and all  derivatives $\d_{(b)}\psi^A\equiv \{\d_{b_1}\ldots \d_{b_k}\psi^A\}$ considered as independent coordinates $\psi^A_{(b)}$. The ghost degree is assigned as follows $\gh{z^a}=0,\gh{\xi^a}=1$ and $\gh{\psi^A_{(b)}}=\gh{\psi^A}$.
On the jet space one defines the total derivative $\d^\tot_a$ by
\begin{equation}
\label{total}
\d^\tot_a=\dl{z^a}+\psi^A_a\dl{\psi^A}+\psi^A_{ab}\dl{\psi^A_b}+\ldots\,.
\end{equation}  
Once the set of fields is chosen all the remaining information, including equations of motion, gauge transformations,
and their (higher) reducibilty relations, is encoded in the BRST differential $s$ which 
is an odd nilpotent vector filed $s$ on the jet space with $\gh{s}=1$. It is assumed evolutionary i.e.
$\commut{\d^\tot_a}{s}=0$. Note that this requirement uniquely determines $s$ in terms of $s^A[\psi]=s\psi^A$.
If the theory is Lagrangian $s$ can be assumed canonically generated by the master action $S_{BV}$ as
$s=\ab{\cdot}{S_{BV}}$, where $\ab{\cdot}{\cdot}$ is the Batalin--Vilkovisky antibracket.

We need to recall the concept of generalized auxiliary fields. Suppose that for the Lagrangian theory
described by the BV master-action $S_{BV}[\psi]$ there is an invertible change
of variables (possibly involving derivatives) $\psi^A\to \phi^i,u^a,u^*_a$ such that 
$u^a,u^*_b$ are conjugate in the antibracket. Fields $u^a,u^*_a$ are said generalized auxiliary
if the equations $\vddl{S_{BV}}{u^a}\big|_{u^*_a=0}=0$ are algebraically solvable with respect to $u^a$.
This notion was introduced in~\cite{Dresse:1990dj}. The reduced master-action is then obtained by substituting
$u^a=u^a[\phi], u^*_a=0$ into $S_{BV}$. Generalized auxiliary fields comprise both usual auxiliary fields and St\"ueckelberg variables as well as their associated ghosts and antifields.

At the level of equations of motion the respective generalization was proposed in~\cite{\BGST}.
According to the definition of~\cite{\BGST} fields $w^a,v^a$ are called generalized auxiliary if there is an invertible change
of variables (possibly involving derivatives) $\psi^A\to \phi^i,v^a,w^a$ such that equations
$sw^a=0, w^a=0$ can be solved algebraically as $v^a=V^a[\phi]$. 
\footnote{More precisely, the prolongation of $sw^a=0, w^a=0$ 
is equivalent to the prolongation of $v^a=V^a[\phi], w^a=0$ as algebraic equations in the jet space.}

Vector field $s$ restricts to the surface $sw^a=0, w^a=0$, giving the reduced theory
whose fields are $\phi^i$ and $s_R\phi^i=(s\phi^i)|_{sw^a=0, w^a=0}$. Following~\cite{\BGST} one shows that
there is an invertible change of coordinates to $\phi^i_R,u^a=sw^a,t^a=w^a$, such that $s\phi^i_R=S_R(\phi_R)$. If in addition this change of coordinates is local (i.e. involves derivatives of finite order only) the generalized auxiliary fields are called local~\cite{Barnich:2010sw}. In this case it is legitimate to rewrite $s$ in the new coordinates as
\begin{equation} 
 s=S^i_R[\phi_R]\dl{\phi^i}+u^a\dl{t^a}+\ldots
\end{equation} 
where dots denote the prolongation. If the number of fields $w^a,v^a$ is finite it is clear that the cohomology of $s$ and $s_R$ in the space of local function(al)s are isomorphic~\cite{\BGST}. This is because the prolongation of the second term
has trivial cohomology and the two pieces do not see each other (the detailed discussion can be found
in~\cite{Brandt:2001tg,\BGST}). The same is true for functional multivectors. For instance, for the cohomology
of $\commut{s}{\cdot}$ in the space of (evolutionary) vector fields~\cite{\BGST}.

If two theories are related through the elimination of auxiliary field of this type 
they are equivalent in a rather strict sense. If the number of auxiliary fields becomes infinite
or the inverse change of variables is not strictly local two theories are usually regarded as equivalent but
in this case the precise equivalence has to be analyzed more carefully. In particular, various isomorphisms
generally hold only for representatives satisfying some extra conditions. Equivalence of this type relates for instance the parametrized and non-parametrized form of the same system or the conventional and the unfolded form. The
parent formulations of~\cite{\BGST,\BGnlp,Grigoriev:2010ic} also belong to this class. The subtleties
arising when the number of auxiliary fields becomes infinite are discussed in some details in Section~\bref{sec:subtle}.

\subsection{AKSZ sigma models}

Consider two $Q$ manifolds, i.e., supermanifolds equipped with an odd
nilpotent vector field~\cite{Schwarz:1992gs}.  The first, called the
base manifold, is denoted by $\manX$. It is equipped with a grading
$\mathrm{gh}_\manX$ and an odd nilpotent vector field $\derham,\,\mathrm{gh}_\manX(\derham)=1$.
We also assume that $\manX$ is equipped with the volume form compatible with $\derham$.  For our purpose
it is enough to restrict to $\manX$ of the form $\Pi T\manX_0$, i.e. to an odd
tangent bundle over a manifold $\manX_0$ which plays a role of space-time.

If $x^\mu$ and $\theta^\mu$ are coordinates on $\manX_0$ and the fibers of $\Pi T\manX_0$ respectively,
the differential and the volume form are given explicitly by
\begin{equation}
 \derham=\theta^\mu \dl{x^\mu}  \,,\qquad  dx^0\dots dx^{n-1}
 d\theta^{n-1}\dots d\theta^0\equiv d^nx d^n\theta\,, \qquad
n=\dim \manX\,.
\end{equation}

The second supermanifold, called the target manifold, 
is denoted by  $\manM$ and is equipped with another degree $\mathrm{gh}_\manM$, the degree $n-1$ symplectic 2-form
$\sigma$ and the function $S_\manM$ with ${\rm gh}_{\manM}(S_\manM)=n$ satisfying the master equation 
\begin{equation}
 \pb{S_\manM}{S_\manM}=0
\end{equation} 
where $\pb{\cdot}{\cdot}$ denotes the (odd) Poisson bracket determined by $\sigma$.
In what follows we always assume that the dimension of $\manM$ is countable and typically equipped with extra structures like e.g. suitable filtration. Although generalizations where either $\sigma$ or $\pb{\cdot}{\cdot}$ can be degenerate are of substantial interest we do not discuss them here. Moreover, we restrict ourselves to the case where $\sigma$ is exact i.e. $\sigma = d\chi$ for some $\chi=\chi_A(\psi)d\Psi^A$, where $\Psi^A$ are local coordinates on $\manM$. 

Given the above data the Batalin-Vilkovisky master-action determining the AKSZ sigma model
is given by
\begin{equation}
\label{AKSZ-action}
 S_{BV}[\Psi,\Lambda]=\int d^nx d^n\theta \left(\chi_A(\Psi(x,\theta))\derham \Psi^A(x,\theta) + S_\manM(\Psi(x,\theta)\right)\,.
\end{equation} 
By construction it satisfies the master equation $\ab{S_{BV}}{S_{BV}}=0$ with respect to the functional antibracket induced by the target space bracket $\pb{\cdot}{\cdot}_\manM$. More precisely, for two functionals $F,G$ one has
(see e.g.~\cite{\GD,Barnich:2009jy} for more details)
\begin{equation}
\label{eq:path-bracket}
\ab{F}{G}=(-1)^{(\p{G}+n)n}
\int d^nx d^n\theta\,
\big(\vddr{F}{\Psi^A(x,\theta)} E^{AB}(\Psi(x,\theta))
\vddl{G}{\Psi^B(x,\theta)}\big)\,,
\end{equation}
where $E^{AB}(\Psi)=\pb{\Psi^A}{\Psi^B}$ are coefficients of the Poisson bivector.

The space of fields is equipped with a total ghost degree $\gh{A}={\rm gh}_{\manM}(A)+ {\rm gh}_{\manX}(A)$.
In particular $\gh{S_{BV}}=0$.
It is useful to identify component fields as 
\begin{equation}
 \Psi^A(x,\theta)=\overset{0}\Psi{}^A(x,\theta)+\theta^\mu\overset{1}\Psi{}_\mu^A(x)+
\half\theta^\mu\theta^\nu\overset{2}\Psi{}_{\nu\mu}^A(x)+\ldots+\frac{1}{n!}\theta^{\mu_n}\ldots\theta^{\mu_1}
\overset{n}\Psi{}_{\mu_1\ldots \mu_n}^A(x)\,.
\end{equation} 
Note that $\gh{\overset{k}\Psi{}^A_{\mu_1\ldots \mu_k}}={\rm gh}_\manM({\Psi^A})-k$.
As usual, fields of vanishing degree are physical ones, those of positive degree are ghost fields while
negative degree ones are antifields. The Lagrangian, generators of gauge symmetries, and higher structures
of gauge algebra are encoded in $S_{BV}$ in a standard way.

The AKSZ sigma model BRST differential is canonicallly generated by $S_{BV}$ canonically to be
\begin{equation}
\label{s-AKSZ}
 s  \Psi^A(x,\theta)=(-1)^n\ab{\Psi^A(x,\theta)}{S_{BV}}=\derham \Psi^A(x,\theta)+Q^A(\Psi^A(x,\theta))\,,
\end{equation} 
where $Q^A(\Psi)=\pb{\Psi^A}{S_\manM}$ and the extra sign-factor $(-1)^n$ has been introduced for
future convenience.

As the BRST differential can be defined just in terms of $\derham$ and an odd nilpotent vector filed $Q$
on the target space one can consider the equations of motion version of AKSZ sigma model.
In this case $\manX$ is the same while on target space one only assumes
existence of a nilpotent vector field $Q, \gh{Q}=1$. This data is enough to equip the space of fields
$\Psi^A(x,\theta)$ with the nilpotent BRST differential using~\eqref{s-AKSZ}.

\noindent
To conclude the discussion of AKSZ sigma models let us discuss their general properties.

-- 
If the space-time dimension 
$n>1$ and the dimension of $\manM$ is finite (so that the number of fields is finite as well)
AKSZ sigma model is necessarily topological. See e.g.~\cite{Ikeda:2012pv} for a recent review of AKSZ approach to topological theories.

-- 
In contrast to the Lagrangian case where the
degree of $S_\manM$ equals the space-time dimension in this case there is no such dependence as ghost degree of $Q$
is always one. In particular, given $\manM$ one can consider a family of models by taking one or another $\manX$.
In particular, this gives a natural way to restrict a gauge system to a submanifold or a boundary~\cite{Vasiliev:2001zy,Gelfond:2003vh,Barnich:2006pc,Bekaert:2009fg} (see also~\cite{Vasiliev:2012vf} for a recent applications in the context of AdS/CFT correspondence).

--  It was observed in~\cite{Barnich:2005ru} that if $\gh{\Psi^A}\geq 0$ for all coordinates then each $\Psi^A$ gives rise to a $p_A$-form physical field with $p_A=\gh{\Psi^A}$ and the equations of motion determined by $s$
take the form of a free differential algebra. In this case the AKSZ sigma model at the level of equation of motion
is simply a BRST extension of the unfolded form. In the general case equations of motion combine both the algebraic constraints and the FDA relations. Moreover,  both the FDA relations and the constraints are encoded in one and the same odd nilpotent vector field $Q$.

-- If one restricts to an appropriate neighborhoods in the space-time and the target space the local BRST cohomology
of the AKSZ sigma model i.e. cohomology of $s$ in the space of local functionals, is isomorphic to the cohomology of the target space differential $Q$ in the space of functions on the target space~\cite{Barnich:2005ru}. If $\manM$ is infinite-dimensional the isomorphism in general requires extra assumptions (see~\cite{Barnich:2010sw} and
the discussion in Section~\bref{sec:subtle}). 

-- Although in the Lagrangian setting AKSZ sigma model is formulated within the Lagrangian BV formalism
the respective Hamiltonian formulation is also built in. More precisely decomposing the space time manifold $\manX_0$
into spatial part $\manX^{s}_0$ with coordinates $x^i$ and the time-line $x^0$ the BRST charge and the
BFV Poisson bracket are given by the same expressions~\eqref{AKSZ-action} and \eqref{eq:path-bracket}
with $\manX_0$ replaced with $\manX_0^{s}$. In particular, the BRST charge reads as
\begin{equation}
\label{brst-charge}
 \Omega=\int d^{n-1}x d^{n-1}\theta (\chi_A(\Psi(x,\theta))\derham_s \Psi^A(x,\theta) + S_\manM(\Psi(x,\theta))\,.
\end{equation} 
where $\derham_{s}=\theta^i\dl{x^i}$ denotes the de Rham differential of $\manX_0^{s}$.
This fact was originally observed in~\cite{Grigoriev:1999qz,Barnich:2003wj}.

-- The previous property can be understood from a more general perspective.
Consider the integrand $L_{AKSZ}$ of the AKSZ action as an inhomogeneous
differential form (in the sense of identification $\theta^\mu\equiv dx^\mu$) on
$\manX_0$ extended by the space of fields and their derivatives. Given a
submanifold $\manZ_0 \subset \manX_0$ of dimension $k$ one can integrate over $\manZ_0$ the $k$-form component of $L_{AKSZ}$. This results in ``higher BRST charges'' of
ghost degree $n-k$. By construction these again satisfy the master equation with
respect to the bracket of degree $k-n+1$ determined by~\eqref{eq:path-bracket}
with $\manX_0$ replaced by $\manZ_0$.  Of a special interest is the case where $\manZ_0$
is a boundary (this was recently discussed in~\cite{Cattaneo:2012qu}).
If $\manZ_0$ is a spatial slice $\manX^{s}_0$,
$k=n-1$ one clearly reproduces~\eqref{brst-charge} and other BFV structures.
Analogous charges~\cite{Vasiliev:2005zu} can be defined by integrating over $\manZ_0$ various $Q$-invariant functions on $\manM$ pulled back by $\Psi^A(x,\theta)$.

-- For an AKSZ sigma model one can identify a special class of generalized
auxiliary fields whose elimination can be performed entirely in the target
space. For as AKSZ model at the level of equation of motion these were proposed
in~\cite{Barnich:2005ru} and are simply coordinates $w^a,v^a$ on $\manM$ such
that $Qw^a=v^a$ can be solved with respect to $v^a$ at $w^a=0$. This is just a
version of a general definition for a $0$-dimensional field theory. In the
Lagrangian setting one again repeats the definition from Section~\bref{sec:af}
treating $\manM$ as a space of fields of $0$-dimensional theory and using
$S_\manM$ in place of the master-action.

\section{Parent formulation}
\label{sec:parent-main}
\subsection{Parent Lagrangian}
\label{sec:parent}

Given a Lagrangian gauge field theory on space-time manifold $\manX_0$ one can embed it into the Batalin-Vilkovisky
description by adding ghost fields and antifields so that the action is promoted to the BV master-action of the form 
\begin{equation}
 S[\psi,\psi^*]=S_0[\psi]+\int d^nx \psi^*_A \gamma \psi^A+\ldots
\end{equation} 
where $S_0$ is the starting point classical action, $\gamma$ the gauge part of the BRST differential, and dots denote
higher order terms in antifields $\psi^*_A$ needed if the gauge algebra does not close off-shell. In what follows we assume
that the algebra is closed and hence $\gamma^2=0$ off-shell so that $S$ can be taken linear in antifields $\psi^*_A$. On the space of fields and antifields there is a canonical odd 1-form $\chi=\psi^*_A d\psi^A$. It gives rise to the canonical antisymplectic structure $\sigma=d\chi=d \psi^*_A\wedge d\psi^A $ and the respective odd Poisson bracket (antibracket).

The parent formulation is constructed as follows: if $z^a$ denote particular space-time coordinates
consider the jet space associated with fields $\psi^A$. This is a supermanifold $\manN$ with coordinates $z^a,\xi^a\equiv dz^a, \psi^A$ and all the derivatives $\psi^A_{(b)}=\d_{(b)}\psi^A$ considered as independent coordinates. 
In addition to the ghost degree $\manN$ is equipped with an odd nilpotent vector field called  the total de Rham differential $d_H=\xi^a \d^\tot_{a}$, where $\d^\tot_{a}$ is a total derivative~\eqref{total}. Furthermore, in these terms
gauge BRST differential $\gamma$ is promoted to an odd nilpotent vector field on $\manN$  satisfying $\commut{d_H}{\gamma}=0$ and $\gh{\gamma}=1$. A useful object which plays an important role in the formalism is the total differential $\tilde\gamma=-d_H+\gamma$. It is nilpotent thanks to $d_H^2=0$, $\gamma^2=0$, and $\commut{d_H}{\gamma}=0$.

As a next step to each coordinate on $\manN$ (collectively denoted by $\Psi^\alpha$) 
one associates a conjugated coordinate $\Lambda_\alpha$ with
$\gh{\Lambda_\alpha}=-\gh{\Psi^\alpha}+n-1$. More precisely one extends $\manN$ to
(odd) cotangent bundle $\manM=T^*[n-1]\manN$  with the degree of the fiber
coordinates shifted by $n-1$. $\manM$ is equipped with a canonical 1-form
$\chi=\Lambda_\alpha d\Psi^\alpha$ of ghost degree $n-1$. This defines a canonical (odd)
Poisson bracket on $\manM$  with  
\begin{equation}
\pb{\Psi^\beta}{\Lambda_\alpha}_{\manM}=\delta_\alpha^\beta\,, \qquad \gh{\pb{\cdot}{\cdot}_\manM}=-n+1\,.
\end{equation} 

The parent formulation is then given by a certain infinite-dimensional version of
the AKSZ sigma model with the target space being $\manM$ and the source space $\Pi T\manX_0$
with coordinates $x^\mu,\theta^\mu$. The target space BRST potential is given by
\begin{equation}
 S_\manM=\Lambda^*_\alpha \tilde\gamma \Psi^\alpha+\hat L(\Psi)\,, \qquad \gh{S_\manM}=n, \quad \p{S_\manM}=n\, \mod \,2
\end{equation} 
where $\hat L$ denotes a representative of the starting point Lagrangian in the cohomology of $\tilde\gamma$, i.e.
if $L^n=\xi^0\ldots\xi^{n-1}L[\Psi,y]$ is a Lagrangian density then $\hat L=L^n+L^{n-1}+\ldots$
is its $\tilde\gamma$-invariant completion by terms of homogeneity $n-1, n-2, \ldots$ in $\xi^a$.
Thanks to nilpotency of $\tilde\gamma$ and $\tilde\gamma\hat L=0$ BRST potential $S_\manM$ satisfies the master equation
$\pb{S_\manM}{S_\manM}_\manM=0$.  Finally, the BV master-action has the standard AKSZ form~\eqref{AKSZ-action}. Explicitly, one has
\begin{equation}
\label{parent-action}
 S_{P}=\int d^nx d^n \theta\left[\Lambda_\alpha(x,\theta)( \derham  +
\tilde\gamma) \Psi^\alpha(x,\theta)+\hat L(\Psi(x,\theta))
\right]\,.
\end{equation} 
It satisfies master equation $\ab{S}{S}=0$ in terms of the antibracket $\ab{\,\cdot\,}{\,\cdot\,}$ 
on the space of fields and antifields, which is determined by the target space bracket $\pb{\cdot}{\cdot}_\manM$ through~\eqref{eq:path-bracket}. Note that thanks to the manifest coordinate-independence of the construction the action is written in terms of generic space-time coordinates $x^\mu$.

Strictly speaking BV master-action~\eqref{parent-action} corresponds to the
parametrized form of the starting point theory. If the starting point theory is
diffeomorphism invariant then one can consistently eliminate component fields
entering $z^a(x,\theta),\xi^a(x,\theta)$ along with their conjugate antifields~\cite{\BGnlp,Grigoriev:2010ic}.
In general, the parent description of non-parametrized theory is achieved by
imposing the gauge condition $z^a(x,\theta)=Z^a(x)$, with $Z^a(x)$ defining an admissible coordinate system.
More precisely, component fields entering $z^a(x,\theta)-Z^a(x),\xi^a(x,\theta)$ and their conjugate antifields are generalized auxiliary and can be eliminated. The elimination simply results in putting
$z^a(x,\theta)=Z^a(x)$ and $\xi^a(x,\theta)=\theta^\mu \d_\mu Z^a$ inside the BV action.

Instead of the minimal $\gamma$ encoding just the gauge symmetry one can start with the non-minimal $\gamma$
and the extended manifold $\manN$. If the extended system is equivalent to the minimal one through the elimination
of generalized auxiliary fields then the respective parent formulations are also equivalent. In particular,
all the considerations including the equivalence proof etc. remain valid in the nonminimal setting as well.
This type of generalization is suitable if, for instance, one wants to take into account off-shell constraints of the starting point theory implicitly through the appropriately extended $\gamma$. In this setting one naturally has negative ghost degree variables among coordinates on $\manN$. Note that in this way one cannot handle differential constraints because the respective Lagrange multipliers (present among components of $\Lambda^\alpha$) can become dynamical.

As the target space is infinite-dimensional the theory is not completely defined
just by specifying the set of fields and the BV master-action even if we
disregard (as we do in any case) the subtle issues of boundary conditions, global geometry etc. 
As we are going to see in the next section it crucially depends on the choice of the space of allowed field
configurations.

A simple way to avoid these subtleties is to truncate the theory by hands to a
finite one as was originally proposed in~\cite{Grigoriev:2010ic}. Another
option we are going to describe next is to keep infinite amount of fields but
specify the space of allowed field configuration in such a way that the parent
formulation is indeed equivalent through the elimination of the generalized
auxiliary fields to the starting point theory without any artificial truncation.

\subsection{Space of allowed field configurations: a simple example}
\label{sec:mech}
The proper definition of the space of allowed configuration is in fact not model specific. 
A good strategy is then to start with the simplest example to illustrate the idea. To this
end consider the parent formulation of a mechanical system with regular Lagrangian $L(q_0,\d q_0)$. 
To keep notations uniform with the field theory generalization we use $\d a$ to denote $\dot a=\frac{d}{dt}a$.\footnote{Although we restrict Lagrangian not to depend on higher order derivatives the formalism is perfectly suited for higher derivative theories. In the case of mechanics it essentially reproduces the familiar Ostrogradsky construction.}

The non-parametrized version of the parent formulation has the following fields: $q_{(l)},p_{(l)}$
$l=0,1,\ldots$ and their conjugate antifields. In this example
antifields are completely passive because they don't enter the BV action and we systematically disregard them. 
Parent action for mechanics is known in the literature and is given by
\begin{equation}
\label{parent-mech}
 S=\int dt \left[p_0(\d q_0 -q_{1})\,+\,\sum_{i=1}^\infty 
p_{i}(\d q_{i}-q_{i+1}) +L(q_0,q_{1})\right]\,.
\end{equation} 

As the space of allowed field configurations let us chose
arbitrary\footnote{The irrelevant in this context choice of smoothness class
and the boundary conditions is not explicitly discussed.}  configurations for
$q_{l}$. As for the configurations for $p_{l}$ we take those where only finite
number of $p_{l}$ are nonvanishing. 

This choice has a simple explanation: $p_{l}$  are coordinates on the space 
dual to the infinite dimensional space with coordinates $q_{l}$. As $q_{l}$ are coordinates on the infinite jet space
and hence an infinite number of $q_{l}$ can be nonvanishing simultaneously the standard choice of the dual space
is to take functionals for which only finite number of components $p_l$ can be nonvanishing.

In this way one reformulates the restriction in terms of the target space rather than a field configuration space. Indeed the target space is a version of a cotangent bundle over the jet space such that it consists of points where only finite number of fiber coordinates (i.e. momenta $p_{l}$) are nonvanishing. In this form this choice is immediately extended to the general parent formulation. Indeed, as the target space of the AKSZ representation is an (odd) cotangent bundle over the BRST extended jet space one again requires that only finite number of fiber coordinates can be nonzero.

Let us show that under the above condition all the variables $q_{i+1}$, $p_i$ with $i=0,1,\ldots$
are auxiliary fields. The equations of motion derived from~\eqref{parent-mech} by varying with respect to
$p_i$, $q_{i}$ with $i=0,1,\ldots$ read as
\begin{equation}
\label{eom-mech}
q_{i+1}= \d{q}_{i} \quad i\geq 0\,, \qquad  p_{i-1}=-\d{p}_{i}+\delta_i^1 \ddl{L}{q_1}+\delta_i^0 \ddl{L}{q_0}
\end{equation} 
where $i=0,1,\ldots$ and $p_i=0$ for $i<0$.

Even without the above condition on the allowed field configurations the first
equation is algebraically solved with respect to $q_{i}$ with $i>0$ by
$q_{i}=(\d)^i q$.  The second equation can not be solved algebraically for $p_i$ 
if all $p_{l}$ can be nonzero. However, in the space where only finite
number of $p_{l}$ can be nonvanishing the second equation with $i>0$ is equivalent to
$p_0=\dl{q_1}L$ and $p_{i}=0$ for $i>0$. One then concludes that for such a field
configuration space variables $p_l,q_{l+1}$ with $l>0$ are auxiliary fields so that
together with their conjugate antifields they are generalized auxiliary fields 
(see Section~\bref{sec:af}). Moreover, their elimination brings back the
starting point system with Lagrangian $L(q,q_{(1)})$. Note that one can choose not to eliminate $p_0$
if $\ddl{L}{q_1}=p_0$ can be solved with respect to $q_1$. This possibility results in the Hamiltonian
formulation of the system (see~\cite{Grigoriev:2010ic} for more details).

\subsection{General case}
In the general case consider a non-parametrized version of the parent formulation which is obtained by imposing the gauge
$z^a=\delta^a_\mu x^\mu$ and hence $\xi^a=\delta^a_\mu \theta^\mu$.  It is convenient to pack
all the fields into the generating functions
\begin{equation}
 \tilde\psi^A=\sum_{p,k\geq 0}\frac{1}{p!k!}\theta^{\mu_p}\ldots\theta^{\mu_1}y^{\nu_k}\ldots y^{\nu_1} \psi^A_{{\nu_1}\ldots {\nu_k}|{\mu_1}\ldots{\mu_p}}\,,
\end{equation} 
where in addition to $\theta^\mu$ we have introduced extra variables $y^\mu$.

In the general case the space of allowed field configuration is specified as follows:
no extra restrictions for $\psi^A_{{\mu_1}\ldots{\mu_p}|{\nu_1}\ldots {\nu_k}}$ but for their conjugate antifields
(i.e. component fields entering $\Lambda^A_{(\nu)}(x,\theta)$) only finite number are allowed to
be nonzero. Note that this reproduces the restriction in the example of the previous section.

Let $f_\alpha, g_\alpha$ be homogeneous monomials in $y,\theta$ such that together with $1$ they form a basis
in polynomials in $y,\theta$ and $\sigma f_\alpha =g_\alpha$ where $\sigma=\theta^\mu\dl{y^\mu}$. We have the following decomposition $\tilde\psi^A=\psi^A+v^A_\alpha f^\alpha+w^A_\alpha g^\alpha$. We use the following collective notations
$v^a$ for $v^A_\alpha$ and $w^a$ for $w^A_\alpha$ so that the entire set of fields is given by $\psi^A,v^a,w^a$
and their conjugate antifields $\psi^*_A,w^*_a,v^*_a$. The parent master-action takes the form (see~\cite{Grigoriev:2010ic}
for more details):
\begin{equation}
 \int d^n x \left[(\int d^n\theta \Lambda^{(\nu)}_A(x,\theta)\derham \psi^A_{(\nu)}(x,\theta))+w^*_av^a+w^*_a\bar\gamma w^a\right]\,,
\end{equation} 
where $\bar\gamma$ denotes the extension of $\gamma$ to component fields entering $\psi^A(x,\theta)$.

Let us show that variables $w^*_a,v^a$ and $w^a,v^*_a$ are generalized auxiliary fields. We need to show that 
equations $\vddl{S_p}{w^*_a}=0$ and $\vddl{S_p}{v_a}=0$ can be algebraically solved for $w^*_a,v^a$ at $w^a=v^*_a=0$.
Under the resolvability we assume resolvability in terms of homogeneity expansions in fields (one can actually require less) so that it is enough to consider the linearized system. The proof that $\vddl{S_p}{w^*_a}=0$ can be solved with respect to $v_a$ 
was given in~\cite{Grigoriev:2010ic}. It is unaffected as the dual variables are not involved in the relevant equations.

As for the $\vddl{S_p}{v_a}=0$ it amounts to showing that $w^*_a-\vdl{v_a}(w^*_b(d^F+\gamma)w^b)|_{w=0}=0$
can be solved for $w^*_a$. Here, $d^F$ is a vector field in the space of fields $\psi^A_{{\mu_1}\ldots{\mu_p}|{\nu_1}\ldots {\nu_k}}$ and their $x$-derivatives that represents the action of $\derham=\theta^\mu\d_\mu$. More precisely,
$d^F$ is defined through  $d^F\tilde\psi^A=\derham \tilde\psi^A$ where $d^F$ and $\d_\mu$ acts on component fields and their derivatives while $\theta^\mu$ entering $\derham$ in the space of auxiliary variables. For instance one gets
$d^F\psi^A_{()|\mu}=(-1)^{\p{A}}\d_\mu \psi^A_{()|[]}$.

Using the degree defined by $\deg{\psi^A_{{\nu_k}\ldots {\nu_k}|{\mu_1}\ldots{\mu_p}}}=p+k-(N+1)\gh{\psi^A}$,
where $N$ is the maximal number of space-time derivatives in $\gamma$, one shows that $\vdl{v_a}(w^*_b(d^F+\gamma)w^b)|_{w=0}$ can only depend on $w^*_c$ of the degree higher than that of $w_*^a$ (by definition, the degree of $w^*_a$ equals the degree of its conjugate $w^a$). Because in a given configuration there can be only finite number of nonvanishing $w_a^*$ variables the equations can be always solved algebraically.

\subsection{Scalar field example}
The parent formulation for a free scalar field in Minkowski space is almost trivial as there is no
genuine gauge symmetry. However, this example is extensively used as an illustration of the unfolded approach as e.g. in~\cite{Shaynkman:2000ts,Kaparulin:2011zz} and at the same time was not discussed in detail in~\cite{Grigoriev:2010ic}.
Moreover, in Section~\bref{sec:lagrange} we need this example to illustrate the relationship to other approaches. 

Scalar field on $n$-dimensional Minkowski space is described by Lagrangian 
$L=-\half \eta^{ab}\d_a\phi \d_b\phi-V(\phi)$ (metric $\eta_{ab}$ is assumed of almost positive signature). The manifold
$\manN$ is the usual jet-space with coordinates $z^a,\xi^a,\phi_{a_1\ldots a_k}, k=0,1,\ldots$.
The target space $\manM$ is the cotangent bundle $T^*[n-1]\manN$ with the shifted degree. Denoting by 
$p_a,\rho_a,\pi^{a_1\ldots a_k}$ the variables conjugate to $z^a,\xi^a,\phi_{a_1\ldots a_k}$
the target space function $S_\manM$ takes the form:
\begin{equation}
 S_M=-p_a \xi^a-\sum_{k=0}^\infty \pi^{a_1\ldots a_k}\xi^a \phi_{aa_1\ldots a_k}
+\Vol\, L\,,\qquad \Vol=\frac{1}{n!}\epsilon_{a_0\ldots a_{n-1}}\xi^{a_0}\ldots \xi^{a_{n-1}}\,.
\end{equation}
The BV master-action~\eqref{parent-action} is given by 
\begin{multline}
S_{P}=\int d^n x d^n \theta\big[ p_a(\derham z^a -\xi^a)+\rho_a\derham \xi^a+\\+\sum_{k=0}^\infty \pi^{a_1\ldots a_k}(\derham \phi_{a_1\ldots a_k}-\xi^a \phi_{aa_1\ldots a_k})+\Vol \,L\big]\,.
\end{multline} 

We then work with the nonparametrized version of the parent formulation which is obtained by putting $z^a=x^a$ and $\xi^a=\theta^a$
inside the BV action which is then understood as a functional of only $\pi^{(a)}(x,\theta)$ and $\phi_{(a)}(x,\theta)$.
Here $x^a$ denote Cartesian space-time coordinates.

Integrating then over $\theta^a$ and keeping only physical (=vanishing ghost degree) fields gives
\begin{equation}
 S_0=\int d^n x \left[\sum_{k=0}^\infty \pi^{b|a_1\ldots a_k}(\d_b\phi_{a_1\ldots a_k}-\phi_{ba_1\ldots a_k})-(\half\phi_a \phi^a+V(\phi))\right]\,,
\end{equation} 
where we have identified $n-1$-forms $\pi^{a_1\ldots a_k}$ with vector fields $\pi^{b|a_l\ldots a_k}$.  Unlike
$\phi_{a_1\ldots a_k}$ fields $\pi^{b|a_1\ldots a_k}$ are subject to gauge invariance which can be read off from
the above master-action. More precisely along with the physical fields $\pi^{b|a_1\ldots a_k}$ target space coordinates
$\pi^{a_1 \ldots a_k}$ give rise to ghost fields which are $n-2$-forms and which we identify with bivectors
$\lambda^{bc|a_1\ldots a_k}$. The gauge transformation reads as
\begin{equation}
\label{scalar-gt}
 \delta_\lambda \pi^{b|a_1\ldots a_k}=\lambda^{b(a_1|a_2\ldots a_k)}-\d_c\lambda^{cb|a_1\ldots a_k}\,,\qquad k=0,1,\ldots
\end{equation} 
where the parenthesis denote the symmetrization of the enclosed indexes. Thanks to the first term this gauge invariance is
St\"ueckelberg (algebraic) for $k>0$ and it is easy to eliminate the respective pure gauge components. 

An efficient way to analyze the system is to introduce suitable Fock space notation. To this end let
$y^a$ be bosonic variables seen as creation operators. The Fock space is then simply the space of polynomials in $y^a$ where
$1$ is the vacuum.
It is also useful to introduce the dual Fock space generated by dual operators $\bar y_a$. The inner product is completely determined by $\inner{1}{1}=1$ and the conjugation rule $(\bar y_a)^*=\dl{y^a}$ and $(y^a)^*=\dl{\bar y_a}$, e.g. $\inner{\bar y_a}{y^b}=\delta_a^b$. With the help of extra fermionic variables $\bar \theta_a$ let us introduce generating functions
\begin{equation}
\Pi(\bar y,\bar\theta)=\sum_{k=0}^\infty \pi^{b|a_1\ldots a_k}\bar \theta_b \bar y_{a_1} \ldots \bar y_{a_k}\,,\quad
\Lambda(\bar y,\bar\theta)=\half\sum_{k=0}^\infty \lambda^{bc|a_1\ldots a_k}\bar\theta_c\bar \theta_b \bar y_{a_1} \ldots \bar y_{a_k}\,,
\end{equation} 
for fields $\pi^{b|a_1\ldots a_k}$ and gauge parameters $\lambda^{bc|a_1\ldots a_k}$. In these terms gauge transformations~\eqref{scalar-gt} can be written as
\begin{equation}
\delta_\lambda \Pi=A\Lambda+ B \Lambda\,, \qquad A= \bar y_a \dl{\bar \theta_a}\,, \quad B=\dl{\bar \theta_a}\dl{x^a}\,,
\end{equation} 
Operator $A$ can be seen as a version of de Rham operator and its cohomology is trivial in nonvanishing degree in $\bar y,\bar \theta$. It follows that by choosing suitable $\Lambda$ one can always achieve $\bar\theta_a\dl{\bar y^a}\Pi=0$ or in other words assume fields $\pi^{b|a_1\ldots a_k}$ symmetric in all indexes.

This motivates the following useful choice of generating functions for $\pi^{b|a_1\ldots a_k}$ after gauge fixing
\begin{equation}
\label{gen-pi}
 \pi_l= \frac{1}{(l+1)!}\pi^{a|a_1\ldots a_l}\bar y_{a}\bar y_{a_1}\ldots \bar y_{a_l}\,.
\end{equation}
It is also useful to introduce generating functions for fields $\phi_{a_1\ldots a_l}$ as
\begin{equation}
\label{gen-phi}
 \phi_l=\phi_{a_1\ldots a_l}y^{a_1}\ldots y^{a_l}\,.
\end{equation}
Here and below the identification $\phi_0\equiv \phi$ is assumed.

In terms of the generating functions the gauge fixed version of the parent action reads as
\begin{equation}
\label{parent-gf-scalar}
 S_0^{gf}=\int d^nx\left[ \sum_{l=0}^\infty \inner{\pi_l}{\d\phi_l-\phi_{l+1}}-\half\phi_a \phi^a-V(\phi)\right]\,,
\end{equation} 
where $\d=y^a\dl{x^a}$. This action is not anymore gauge invariant and is clearly a direct generalization of the 
1d action~\eqref{parent-mech}. It can be seen as a minimal local action whose set of fields contains fields of the unfolded formulation and whose equations of motion reproduce the unfolded ones.  Another proposal for the unfolded equations variatinal principle was pushed forward inin~\cite{Shaynkman:2000ts}.

The equations of motion obtained by varying~\eqref{parent-gf-scalar} with respect to $\pi_l$ and $\phi_l$
read as
\begin{equation}
\label{eom-scalar}
 \d\phi_l-\phi_{l+1}=0\,,\qquad \bar \d \pi_l+\pi_{l-1}+\delta_l^1\phi_1(\bar y)+\delta^0_l\ddl{V}{\phi}=0\,, \qquad l=0,1,\ldots\,,
\end{equation} 
where $\bar \d =\dl{\bar y_a}\dl{x^a}$, $\phi_1(\bar y)=\phi_a\eta^{ab}\bar y_b$,
and $\pi_{i}=0$ for $i<0$.

Just like in the case of mechanics if only finite number of $\pi_l$ can
be nonzero then equations~\eqref{eom-scalar} can be solved for $\phi_l,l> 0$ and $\pi_l, l\geq
0$ algebraically so that these variables are auxiliary fields. Indeed, equations
obtained by varying with respect to these variables can be solved for them
algebraically. The reduced action is obviously the starting point one in agreement
with the general equivalence statement. Note that one can consider a different
reduction (see~\cite{Grigoriev:2010ic} for more details) where one first eliminates $\phi_l, l>1$ and $\pi_l, l>0$ and then eliminates
$\phi_1$ through its own equation of motion. This results in the well-known
first order Schwinger action depending on $\phi$ and $\pi^a$. 

\section{Structure of generalized auxiliary fields of parent formulations}
\label{sec:structure}

\subsection{Relation to the Lagrange structure for the unfolded equations}
\label{sec:lagrange}

For a Lagrangian theory the parent formulation can be considered as a Lagrangian extension of the parent formulation
at the level of equations of motion which in turn is an extension of the unfolded
formalism~\cite{Vasiliev:1988xc,Vasiliev:1988sa,Vasiliev:2003ev}. In our setting it is easy to describe both 
off-shell and on-shell unfolded formulation using the BRST theory terms.

For the off-shell version let us restrict to the case where all coordinates on $\manN$ have nonnegative ghost degree. 
Using the conventions of section~\bref{sec:parent} the set of fields is given by the coordinates on $\manN$, with the
coordinate of ghost degree $p$ giving rise to a $p$-form field, $p-1$-form ghost field, $p-2$-form ghost for ghost etc. Note that no antifields/momenta $\Lambda^\alpha$ are introduced in contrast to the parent formulation of section~\bref{sec:parent}. 
The BRST differential is determined by~\cite{Barnich:2010sw}
\begin{equation}
 s^P\Psi^\alpha(x,\theta)=\derham \Psi^\alpha+\tilde\gamma \Psi^\alpha(x,\theta)
\end{equation} 
or in the appropriate jet-space notations of~\cite{\BGnlp} it can be written as $s^P=d^F+{\tilde\gamma}$. Here by a slight abuse of notation $\tilde\gamma$ denotes the natural extension of $\tilde\gamma$ to component fields entering $\Psi^\alpha(x,\theta)$. It is easy to see that the equations of motion determined by $s^P$ have the form of a free differential algebra.

The on-shell version is arrived at by imposing in addition the prolongation of
the starting point equations of motion understood as constraints in $\manN$. In
more geometric terms this simply amounts to replacing $\manN$ with $\Sigma
\subset \manN$ singled out by the prolongation of equations of motion. Note that
gauge invariance of the equations of motion implies that $\tilde\gamma$ is
tangent to $\Sigma$ and hence restricts to $\Sigma$. In this way one ends up
with the on-shell unfolded system whose equations of motion are equivalent to
those of the starting point theory.~\footnote{However, as the number of fields
is infinite the equivalence is to be understood with some care. See the
discussion in section~\bref{sec:subtle}. Strictly speaking the parametrization
also breaks strictly local equivalence (see the discussion
in~\cite{Barnich:2010sw}) but this can be avoided by using the non-parametrized
version of the parent formulation} The following comments are in order: 

 -- What we have just described is (in general) the parametrized version of
the unfolded system. As originally proposed in~\cite{Vasiliev:1988xc,Vasiliev:1988sa,Vasiliev:2003ev} the unfolded
formulations for field systems on a given gravity background is based on
gauging the space-time symmetry algebra rather than parametrization. Although
the difference is essential for some specific issues it can well be ignored in the present context.
Moreover,  both formulations coincide if one fixes the diffeomorphism invariance in a suitable way
(see~\cite{Barnich:2010sw} fore more details).

-- Although in the case of theories without gauge freedom the formulation
based on above $s^P$ and the usual unfolded formulation coincide (modulo the
above subtlety with parametrization) this is not the case for general gauge
systems. More precisely, the standard unfolded formulation known in the literature
(e.g. of the free spin $s$ gauge field~\cite{Lopatin:1988hz}) is achieved by eliminating a maximal amount of contractible pairs for $\tilde\gamma$ as was originally described in~\cite{Barnich:2004cr,Barnich:2006pc} in the case of free systems (see also~\cite{Vasiliev:2005zu} for particular nonlinear systems). The general
case was described in~\cite{Barnich:2010sw}. 

\bigskip

From the above discussion it is clear that the parent Lagrangian formulation can be
seen as an on-shell unfolded system extended by generalized auxiliary fields in such a way
that the resulting system has the form of a Lagrangian AKSZ sigma model. Alternatively, it can be arrived at by
extending the off-shell unfolded system followed by deforming the resulting AKSZ Lagrangian by $\hat L$.
In this sense the parent Lagrangian formulation gives a Lagrangian Batalin--Vilkovisky extension of the unfolded approach and hence provides a setup for quantization, studying consistent deformations, etc. within a well-established and powerful framework~\cite{Batalin:1981jr,Batalin:1983jr,Barnich:1993vg,Barnich:2000zw}.

Another approach to quantizing unfolded dynamics is based on using so-called
Lagrange structure~\cite{Kazinski:2005eb} instead of a genuine
Lagrangian. In a recent work~\cite{Kaparulin:2011zz} the Lagrange structure for the usual
unfolded form of a free scalar field was constructed. Now we are going to show
that the Lagrange structure of~\cite{Kaparulin:2011zz} can be obtained by
reducing the canonical Lagrange structure of the Lagrangian parent formulation
of~\cite{Grigoriev:2010ic}. Moreover, this gives a systematic way to obtain the
explicit form of the natural Lagrange structure for the unfolded formulation of
a given Lagrangian system. 

We begin with a very brief reminder of the Lagrange structure concept. Details
can be found in the original
papers~\cite{Kazinski:2005eb,Lyakhovich:2005mk,Lyakhovich:2007cw,
Lyakhovich:2004xd} (see also~\cite{Barnich:2009jy}). Given a system of (gauge
invariant) differential equations the Lagrange structure can be introduced
within the BV-BRST framework in the following way: at the level of equations of
motion the gauge system is described by the nilpotent BRST differential. If the
system is a genuine Lagrangian gauge theory one can formulate it in such a way
that the BRST differential is canonically generated in the odd Poisson bracket
(BV antibracket).  The respective generator, the BV master-action, is essentially
the Lagrangian. Moreover, up to technical details specifying BRST-invariant
nondegenerate bracket uniquely fixes the Lagrangian so that the data of the BRST
differential and the compatible bracket can be used to define the system.

In this framework one can allow for not necessarily nondegenerate or regular antibracket
(in fact the Jacobi identity can be also fulfilled in a weak sense only) so the
BRST differential is not anymore canonically generated. The Lagrange structure
is roughly a BRST-invariant antibracket which is allowed to be degenerate. This concept was introduced
in~\cite{Kazinski:2005eb} where it was shown that this data is enough to define
a consistent quantization of the system. This has a simple counterpart in the
Hamiltonian quantization: given a possibly degenerate or not necessarily regular
Poisson bracket one can consistently define its deformation
quantization~\cite{Kontsevich:1997vb}.

Given a nilpotent BV-BRST differential $s\psi^A=s^A[\psi]$ associated to gauge invariant equations of motion one can associate
an odd BRST charge $\Omega_0=\bar \psi_A s^A $ on the space of the field variables $\psi^A$ and their canonically conjugate momenta $\bar\psi_A$. This satisfies the master equation $\pb{\Omega_0}{\Omega_0}=0$ where $\pb{\,}{\,}$ denotes the canonical Poisson bracket determined by $\pb{\psi^A}{\bar\psi_B}=\delta^A_B$. In the case of Lagrangian gauge systems we are interested in now these structures were introduced in~\cite{Grigoriev:1999qz}. For instance for the scalar field described by ~\eqref{parent-gf-scalar} BRST charge $\Omega_0$ takes the form (here and below space-time integrals are implicit)
\begin{multline}
\label{Omega0}
 \Omega_0=
\bar\phi^*_l\vddl{S_0}{\phi_l}+\bar\pi^*_l\vddl{S_0}{\pi_l}=\\
=\sum_{l=0}^\infty \left[ \inner{\bar\pi^*_l}{\d\phi_l-\phi_{l+1}}
-\inner{\bar\phi^*_l}{\bar\d \pi_l+\pi_{l-1}+\delta_l^1\phi_1(\bar y)+\delta^0_l\ddl{V}{\phi}}
\right]\,,
\end{multline} 
where $\phi_{l},\pi_{l}$ were introduced in \eqref{gen-pi}, \eqref{gen-phi}, $\phi^*_l(\bar y),\pi^*_l(y)$ are analogous generating functions for their conjugate antifields, and $\bar\phi^{l}(\bar y),\bar\pi^{l}(y)$ and $\bar\phi^*_l(y),\bar\pi^*_l(\bar y)$ denote generating functions for momenta conjugate to $\phi_{l},\pi_{l},\phi^*_l,\pi^*_l$.
Note that $\gh{\bar\phi^*_l}=\gh{\bar\pi^*_l}=1$ so that $\Omega_0$ has a simple meaning of the BRST charge implementing the equations of motion as Hamiltonian first-class constraints.

The Lagrange structure can be understood as a deformation of linear in momenta $\Omega_0$ by the terms quadratic in
momenta~\cite{Kazinski:2005eb}. The canonical Lagrange structure for a Lagrangian theory is given by
 $\half(-1)^{\p{B}}\bar\psi_A E^{AB} \bar\psi_B$, where $E^{AB}=\ab{\psi^A}{\psi^B}$ are coefficients of the Poisson bivector of the antibracket, $\p{A}=\p{\psi^A}$, and $\bar\psi_A$ denotes momenta conjugate to $\psi^A$. In the case at hand one gets
\begin{equation}
\label{Omega}
 \Omega=\Omega_0+\Omega_1=\Omega_0-\sum_{l=0}^\infty \left[\inner{\bar\pi^*_l}{\bar\pi_l}+\inner{\bar\phi^*_l}{\bar\phi_l}\right]\,,
\end{equation}
with $\Omega_0$ given by~\eqref{Omega0}.
The last two terms are simply the ones encoding the canonical antibracket.

It turns out that the BRST charge $\Omega$ can be used to reduce the theory by
eliminating the generalized auxiliary fields at the level of equations of
motion~\cite{\BGST} pretty much the same way as the master-action can be used to
eliminate generalized auxiliary fields in the Lagrangian
setting~\cite{Dresse:1990dj}. Indeed one can check that replacing the master
action with $\Omega$ and antifields with momenta in the original definition
of~\cite{Dresse:1990dj} describes elimination of the generalized auxiliary
fields at the level of equations of motion along with their conjugate momenta. This
gives an efficient way to reduce both the BRST differential and the Lagrange
structure in a consistent way. As an illustration it is easy to observe that all
the variables entering~\eqref{Omega} save for $\phi_0,\phi^*_0,\bar \phi_0,\bar
\phi_0^*$ are generalized auxiliary and their elimination gives back usual
$\Omega_{\red}=\bar \phi_0^*(\d_a\d^a \phi_0-\ddl{V}{\phi_0})-\bar\phi^*_0\bar\phi_0$
associated to the usual formulation of the scalar field.

Let us consider now a different reduction of the system described by~\eqref{Omega}.  Varying $\Omega$ with respect to
$\pi_l, l\geq 0$ and  $\bar \phi^*_l, l>0$ and  putting  $\phi^*_l, l>0$ and $\bar \pi_l, l\geq 0$ to zero one gets
\begin{equation}
\bar\phi^*_{l+1}=\d\bar\phi^*_l\,,
\qquad 
-\bar\phi_l=\bar\d \pi_l+\pi_{l-1}+\delta_l^1\phi_1(\bar y)
\,.
\end{equation} 
The first equation immediately gives $\bar \phi^*_l=(\d)^l \bar\phi^*_0$. The second equation is solved by
\begin{equation}
 \pi_l=-\sum_{i=0}^\infty (-\d)^i\bar \phi_{l+i+1}-\phi_1\delta^l_0\,.
\end{equation} 
so that they are algebraically solved for  $\bar \phi^*_l, l>0$ and $\pi_l, l\geq 0$
and hence these variables and their conjugate momenta are generalized auxiliary fields.

Substituting these variables in terms of the remaining ones in $\Omega$ gives
 \begin{equation}
\Omega_{red}=
\inner{\bar\pi^*_l}{\d \phi_l-\phi_{l+1}}
+\inner{\bar\phi_0^*}{\bar\d\phi_1(\bar y)-\bar\phi_0-\sum_{i=1}^\infty (-\bar\d)^{i}\bar \phi_{i}}\,.                                                                                                        \end{equation} 
Note that $\bar \d \phi_1(\bar y)=\d^a \phi_a$  and the constraint
$\d\phi=\phi_1$ imply $\bar \d \phi_1(\bar y)=tr \phi_2=\phi_{a}^a$. One then
represents the second parenthesis as
$\phi_a^a-\sum_{i=0}^\infty (-\d)^{i}\bar\phi_{i}$. 
In this form it obviously coincides with the Lagrange structure
extension of the unfolded constraint $\phi^a_a=0$ proposed in~\cite{Kaparulin:2011zz}.
Note that the reduction also reproduces the trivial
Lagrange structure extension of the equations $\d \phi_l-\phi_{l+1}=0$ of this reference. 
Let us recall that in our notation the unfolded form of the scalar field read as
\begin{equation}
 \d \phi_l-\phi_{l+1}=0\,, \qquad tr\phi_l=0\,.
\end{equation}

Let us stress that the Lagrange structure in this setting is represented by a function depending
on derivatives of unbounded order. This type of functions is normally excluded in the usual BRST cohomology treatment.
In particular, this explains that there is no contradiction between the form of the Lagrange structure
and the result of~\cite{Barnich:2009jy} stating that any Lagrange structure for the AKSZ sigma model (and in particular unfolded system) is equivalent to the one not involving space-time derivatives as the considerations of~\cite{Barnich:2009jy}
were explicitly restricted to local function(al)s and hence local Lagrange structures.

\subsection{Functional multivectors and generalized auxiliary fields}
\label{sec:subtle}

The example of the previous section can in fact be easily understood from a more general
perspective. To this end in the setting of Section~\bref{sec:af} let us consider the case where the number
of fields $v^a,w^a$ is infinite. 

As in Section~\bref{sec:af} it is convenient to use adapted coordinates $\phi^i_R,u^a=sw^a,t^a=w^a$
such that $s\phi^i_R=S^i_R[\phi_R]$. Even if the change of variables $\phi^i,w^a,v^a\to \phi^i_R,u^a,t^a$
is local and invertible the inverse change of variables can have unbounded order of derivatives even if the expression
for an individual $v^a$ in terms of $\phi_R,u,t$ is a local function (i.e. contains derivatives of finite order only)
for any given $v^a$. If $f(\phi,v,w)$ is a representative of $s$-cocycle (modulo total derivative) then by re-expressing
it in terms of adapted coordinates as $f^\prime(\phi_R,u,t)$ it is easy to see that $f^\prime(\phi_R,0,0)$
is an $s_R$ cocycle. However, it can involve derivatives of unbounded order if the number of
$v^a$-variables is infinite. That is why the isomorphism of cohomology takes place only if one restricts to local function(al)s depending on finite number of fields only.

This type of generalized auxiliary fields is exactly what one employs to reformulate the theory in the
unfolded form or parent form~\cite{Barnich:2010sw}.~\footnote{Note that if the theory at hand is not diffeomorphism-invariant one needs to chose a particular background or use
the non-parametrized version for the parent formulation. Otherwise
one spoils the strict equivalence by extra gauge fields whose elimination is not a strictly local operation (see \cite{Barnich:2010sw} for more details).} In particular, it was shown in~\cite{Barnich:2010sw} that cohomology in the space of local function(al)s of a given theory and its parent extension are isomorphic if one restricts to function(al)s depending on finite number of fields only. 

In the same setting let us now consider the local BRST cohomology in the space
of functional multivectors. If one restricts to graded symmetric multivectors a
standard way to treat them (see e.g.~\cite{Barnich:2009jy} for more details) is
to introduce momenta conjugate to each variable and then identify functional
multivectors with local functionals homogeneous in momenta. In our case we
introduce momenta $\bar\phi^i,\bar v_a,\bar w_a$ conjugate to  $\phi^i,v^a,w^a$.
To study the relation between the cohomology it is again useful to utilize an
adapted coordinate system $\phi^R,u^a,t^a$, where $u^a=sw^a, t^a=w^a$ and $\phi^i_R$ are complementary coordinates,
along with their conjugate momenta $\bar\phi^i_R,\bar u_a,\bar t_a$.

The change of coordinates $\phi^i,v^a,w^a\to \phi^i_R,u^a,t^a$ now extends to a canonical transformation
$\phi^i,v^a,w^a,\bar\phi^i,\bar v_a,\bar w_a\to \phi^i_R,u^a,t^a,\bar\phi_i^R,\bar u_a,\bar t_a$. If
the inverse change is given by $\phi^i=\phi^i[\phi_R,u,t],v^a=v^a[\phi_R,u,t],w^a=w^a[\phi_R,u,t]$ then the standard momenta transformation law for e.g. $\bar\phi_i$
\begin{equation}
\label{mom-tr}
 \bar\phi^R_i(x)=\int d^n y\left(\bar\phi_j(y)\vddl{\phi^j(y)}{\phi_R^i(x)}+\bar v_a(y)\vddl{v^a(y)}{\phi_R^i(x)}
+\bar w_a(y)\vddl{w^a(y)}{\phi_R^i(x)}\right)\,,
\end{equation} 
where we have used the usual field-theoretical language to simplify the exposition and to avoid introduction
jet-space technique for the momenta. It is clear from the expression that if the number of $v^a$
is infinite the second term may contain derivatives of arbitrarily high order.  Note that the third term vanishes in the present case as $w^a=t^a$.

If now $f(\phi_R,\bar\phi_R,u,t,\bar u,\bar t)$ is an $s$-cocycle expressed in terms of adapted coordinates then $f(\phi_R,\bar\phi_R,0,0,0,0)$ is an $s_R$-cocycle.  Other way around 
if $f(\phi,\bar\phi_R)$ is an $s_R$-cocycle of the reduced theory then considered as a functional in the unreduced formulation it represents $s$-cocycle. Although this map is an isomorphism of cohomology in the space local functionals in $\phi_R,\bar\phi_R,u,t,\bar u,\bar t$, the change of variables can produce derivatives
of unbounded order so that this does not in general induce isomorphism between the original and the reduced system.
Nevertheless, it may happen that such a representative is equivalent to a genuine local one
but this is not guaranteed. This is usually the case with functional vector fields (and hence $f$ linear
in momenta) because one can remove all the derivatives of the momenta by adding total derivatives and then reexpress derivatives of fields through the equations of motion.

This is exactly what happens in the case of parent formulation at the level of equations of motion~\cite{Barnich:2010sw}. Indeed, the parametrized parent formulation is simply an AKSZ sigma model with the target space being the jet space of BRST formulation of the starting point theory.  If $W$ is 
an $s$-cocycle i.e. the evolutionary vector field on the jet space of the starting point theory. By the isomorphism it is mapped to a functional vector field induced by $W$ understood as a vector field on the target space. It is easy to check that in this way one indeed gets a BRST cocycle of the parent formulation. In particular, the above observation shows that it is legitimate to analyze global symmetries within the parent formulation in agreement with~\cite{\BGST,Bekaert:2009fg}.

Let us illustrate the above general considerations using  the following toy model of the unfolded or parent extension. We take as $\phi^i$ variables $\phi_0,\phi_0^*$ (coordinate and its antifield) and as $v^a,w^a$
variables $\phi_l,w_l, l=1,\ldots$ with $\gh{\phi_l}=0$ and $\gh{w_l}=-1$. The BRST differential is
\begin{equation}
 s \phi=s \phi^*=0\,,\qquad s w_l=\phi_l-\d\phi_{l-1}\,,
\end{equation} 
so that we are indeed dealing with the off-shell unfolded system.

Let $f(\phi_0,\phi_0^*,\phi_l,w_l)$ be a representative of
$s$-cohomology. We then reduce the system by eliminating auxiliary fields $\phi_l,w_l, l=1\ldots$. As explained above
we switch to new coordinate system 
\begin{equation}
\phi_R=\phi_0,\quad \phi_R^*=\phi_0^*,\quad u_l=\phi_l-\d\phi_{l-1},\quad t_l=w_l\,.
\end{equation}
  The inverse transformation reads as
\begin{equation}
 \phi_l=u_l+\sum_{i=1}^{l-1}(\d)^i u_{l-i}+(\d)^l \phi_R\,,\quad \phi^*=\phi_R^* \,, \quad w_l=t_l\,.
\end{equation} 
In terms of new variables the map of representatives amounts to simply putting $u_l,w_l$ to zero. In particular $\phi_l$
is mapped to $(\d)^l\phi_R$ as one expects from the very beginning. Note that in this example it is also 
clear that if $f$ depends on infinite number of $\phi_l$ it is mapped to a nonlocal function of $\phi_R$.

We then consider bivectors. As a characteristic example take $\Omega_1=\bar\phi_R\bar\phi_{R}^*$
which encodes a canonical antibracket between $\phi_R$ and $\phi_R^*$ of the reduced system.
In the adapted coordinates $\Omega_1$ represents a cocycle of the extended system as well.
Let us rewrite it in terms of original coordinates: $\bar\phi_{R}^*=\bar\phi_0^*$
while for $\bar\phi_R$ equation \eqref{mom-tr} implies 
\begin{equation}
\bar\phi^R=\sum_{l=0}^\infty \int d^ny \bar\phi_l(y)\vddl{\phi_l(y)}{\phi_R(x)}=\sum_{l=0}^\infty (-\d)^l \bar \phi_l\,,
\end{equation} 
so that $\Omega_1=\sum_{l=0}^\infty\bar\phi_0^* (\d)^l \bar \phi_l$. It should not be a surprise that this
is exactly the Lagrange structure for an unfolded  scalar field. In the previous section we have seen
how to get it from the Lagrangian parent formulation through the appropriate reduction. Now we have derived
it by explicitly adding generalized auxiliary fields.

It is clear from the above discussion that the set of generalized auxiliary fields employed in the Lagrangian parent formulation is very special. Indeed this set contains the subset of fields employed in off-shell unfolded form
(more precisely, the off-shell parent formulation at the equations of motion level) but at the same time the extended Lagrange structure is still canonical. Moreover, auxiliary fields employed in the Lagrangian parent
formulation are not of the type considered above. It is instructive to illustrate this 
using the example of Section~\bref{sec:mech}. 

For the parent formulation of mechanics the structure of auxiliary fields is clear from  equations~\eqref{eom-mech}. Let us try to find a coordinate change 
\begin{equation}
q_{l+1},p_l ~~\to~~ r_{l+1}[q],\,t_l[p]\qquad   l\geq 0\,,
\end{equation}
that brings the auxiliary field equations to the standard form
$r_{l+1}=0,t_l=0$,   $l\geq 0$. For $r$-variables we take $r_l=q_l-\d q_{l-1}$ while for $t$-variables we take $t_l=p_l+\d p_{l+1}-\delta_l^0\ddl{L}{q_1}$. 
The inverse change of variables involves derivatives of arbitrary order. For example, for $p_l$ one gets 
$p_l=\sum_{i=0}^\infty (-\d)^i t_{l+i}+\delta_l^0\ddl{L}{q_1}$. Without the restriction on allowed configuration (so that only finite number of $p_l$ and hence $t_l$ can be nonvanishing) this expression is not even well-defined. 

The analysis of this section (see also the discussion of generalized auxiliary fields in~\cite{Barnich:2010sw} and~\cite{Brandt:2001tg,Kaparulin:2011zz}) shows that the notion of equivalence between theories
differing through  elimination of infinite number of generalized auxiliary fields is somewhat subtle.
The strict equivalence often requires extra requirements on the class of allowed functionals or even
fails in a naive sense. This gives a wide range of possibilities to deform the theory by using its parent formulation. 
We plan to return to this issues elsewhere.

\section{Parent formulations for totally symmetric fields}
\label{sec:HS}

\subsection{Parent formulation at the level of equation of motion}
\label{sec:F-eom}
The free theory of totally symmetric spin-$s$ gauge field is formulated as follows. The set of fields
is given by $\phi^{a_1\ldots a_s}$ which is assumed double-traceless. The ghost field associated to gauge transformations is $C^{a_1\ldots a_{s-1}}$ and is assumed traceless. The contraction of indexes is defined using the Minkowski space metric.
The gauge part of the BRST differential is given by
\begin{equation}
\label{gamma0}
 \gamma \phi= p^a\d_a C\,,
\end{equation} 
where generating functions $\phi,C$ are introduced as follows
\begin{equation}
\phi= \frac{1}{s!}p_{a_1}\ldots p_{a_s}\phi^{a_1\ldots a_s}\,, \quad 
C=\frac{1}{(s-1)!}p_{a_1}\ldots p_{a_{s-1}}C^{a_1\ldots a_{s-1}}\,.
\end{equation}

The Lagrangian is given by~\cite{Fronsdal:1978rb}
\begin{multline}
\label{FL}
 L=\half\inner{\d_a\phi}{\d^a\phi}-\half\inner{\bar p^a \d_a\phi}{\bar p^a \d_b \phi}+\inner{p_a \d^a D}{\bar p_b \d^b\phi}- \\-\inner{\d_a D}{\d^a D}-\half\inner{\bar p^a \d_a D}{\bar p^b \d_b D},
\end{multline}
where $\bar p^a\equiv\dl{p_a}$, $D\equiv T \phi$, and $T\equiv \dl{p_a}\dl{p^a}$. Note the
transformation for $D$: $\gamma D=\bar p^a \d_aC$.

By considering $D$ as an
independent field one can remove (double-) tracelessness condition on fields and
parameters. This essentially coincides with the so-called triplet
formulation~\cite{{Ouvry:1986dv,Bengtsson:1986ys,Henneaux:1987cp}} (see
also~\cite{Pashnev:1998ti,Francia:2002pt,Sagnotti:2003qa,Bekaert:2003uc} for
more recent developments) and the above Lagrangian describes the reducible
system in this case. The irreducible system in the triplet approach is singled
out by the constraints $D=T \phi$, $TD=0$, and $T C=0$. 

To construct parent formulation we introduce supermanifold $\manN$ (jet-space) with
coordinates $z^a,\xi^a$ along with $\phi,C$ and their
derivatives. As usual it is convenient to handle derivatives by allowing $\phi$
and $C$ to depend on extra variables $y^a$
\begin{equation}
\begin{gathered}
\phi(p,y)=\phi(p)+\phi_a(p)y^a+\half \phi_{ab}(p) y^a y^b+\ldots\,, \\
C(p,y)=C(p)+C_a(p)y^a+\half C_{ab}(p) y^a y^b+\ldots\,.
\end{gathered}
\end{equation} 
It is also convenient to introduce the following operators in the space of auxiliary variables $p,y$
\begin{equation}
\sd=p^a \dl{y^a}\,, \quad S=\dl{y^a}\dl{p_a}\,, 
\quad \Box=\dl{y^a}\dl{y^a}\,.
 \end{equation}

The off-shell parent formulation~\cite{\BGST,Barnich:2010sw} is an AKSZ sigma model with the target space $\manN$ equipped with the differential $\tilde\gamma=-d_H+\gamma$. Fields are 1-form $A$ and $0$-form $F$ which are the following component fields
\begin{equation}
\begin{aligned}
 C(x,\theta|y,p)&=\overset{0}C(x|y,p)+\theta^a A_a(x|y,p)+\ldots\,,\\
\phi(x,\theta|y,p)&=F(x|y,p)+\theta^a \overset{1}\phi(x|y,p)+\ldots\,,
\end{aligned}
\end{equation}   
while the component $\overset{0}C(x|y,p)$ has ghost degree $1$ and is a ghost field associated to a parameter of the gauge transformations. 

The equations of motion and gauge transformations are given respectively by~\cite{\BGST}
\begin{equation}
\label{eom-P}
(\derham-\sigma)A=0 \qquad (\derham-\sigma)F+\sd A=0\,,
\end{equation} 
and
\begin{equation}
\label{gs-P}
 \delta A=(\derham-\sigma)\lambda
,\qquad  \delta F=\sd \lambda\,,
\end{equation} 
where $\sigma=\xi^a\dl{y^a}$ and $\lambda=\lambda(x|y,p)$ has the same structure
as $\overset{0}C(x|y,p)$ (i.e. $p^a\dl{p^a}\lambda=(s-1)\lambda$ and
$T\lambda=0$). In addition one has equation and gauge symmetries for fields
implementing reparametrization invariance: $\derham
\overset{0}z{}^a=\overset{1}\xi{}^a$  and $\delta \overset{0}z{}^a=\epsilon^a\,,
\delta\overset{1} \xi{}^a=\derham \epsilon^a$. Recall that at any moment one can
fix reparametrization invariance by e.g. putting everywhere $z^a=x^a$ and
$\xi^a=\theta^a$. 

The on-shell version of the parent formulation is obtained by requiring both $C,\phi$ and hence $A,F$ to be totally traceless: $TC=SC=\Box C=T\phi=S\phi=\Box \phi=0$. It is clear that the equations of motion~\eqref{eom-P} and gauge transformations~\eqref{gs-P} are consistent with the constraints.

Let us now recall the cohomological results of~\cite{\BGST} and demonstrate how
the unfolded formulation can be arrived at in this framework. According
to~\cite{\BGST} (see also \cite{Bekaert:2005ka}) all the fields, ghosts, and
their independent derivatives can be replaced by trivial pairs for
$\tilde\gamma$ except for so called-generalized connections and generalized
curvatures (in the context of general gauge theories these structures were push
forward in~\cite{Brandt:1996mh,Brandt:1997iu}). These can be conveniently packed
into the generating functions $\cC(y,p)$, $R(y,p)$ satisfying\footnote{Note that
if one considers $\cC$ as a 1-form these variables give a set of fields
for the unfolded formulation of Fronsdal system~\cite{Lopatin:1988hz}.} 
\begin{equation}
\label{AFcond}
p^a\dl{y^a}\cC=0\,, \qquad y^a\dl{p^a}R=0\,,
\end{equation}
along with the tracelessness conditions
\begin{equation}
TR=SR=\Box R=0\,,\qquad T\cC=S\cC=\Box \cC=0\,.
\end{equation} 
Recall also the spin conditions for these variables $p^a\dl{p^a} R=sR\,,$ $p^a\dl{p^a} \cC=(s-1)\cC$.

The reduced differential $\tilde\gamma_\red$ is conveniently represented as~\cite{\BGST}
\begin{equation}
 \tilde\gamma_\red \cC=\sigma \cC+\Pi \sigma\bar\sigma R\,,\qquad 
\tilde\gamma_\red R=\Pi \sigma R\,,
\end{equation} 
where $\sigma=\xi^a\dl{y^a}$, $\bar\sigma=\xi^a\dl{p^a}$ while $\Pi$ and $\cP$
denote projectors to the subspaces determined by $\sd \chi=0$ and $y^a \bar p_a
\chi=0$ respectively. In particular, the AKSZ sigma model with the target space
with coordinates $C,R,z,\xi$ and differential $\tilde\gamma_{\red}$ is precisely the
parametrized version of the unfolded formulation~\cite{Lopatin:1988hz} of the Fronsdal
system. 

In what follows we also need the off-shell version of the above reduction. 
If one does not restrict to the stationary surface the reduced set of variables involves~\cite{Bekaert:2005ka} generalized connections encoded in $\cC$, off-shell curvatures $\hat R$ (which are not traceless anymore) and the Fronsdal tensor $\cF$ along with its derivatives. In terms of the fields and their derivatives $\cF$ is expressed as follows
\begin{equation}
\cF=(\Box\phi-\sd S \phi+\sd\sd T)\phi|_{y=0}\,.
\end{equation} 
It follows from $\cF\sd=\sd\sd\sd T$ that $\cF$ is gauge invariant. Indeed, $\gamma \cF \phi =\cF\sd C=
(\sd)^3T C=0$ because $C$ is traceless.

Upon eliminating the contractible pairs the differential $\tilde\gamma=-d_H+\gamma$ becomes a nilpotent ghost degree
$1$ vector field $Q$ on the reduced space. To the best of our knowledge the explicit expression for $Q$ is not available in the literature. However, we need only few explicit relations which can be obtained directly:
\begin{equation}
Q \cC=-\xi^b \cC_b\,,\qquad 
Q \cC_a=-\xi^b\cC_{ba}+
\xi_a\xi_c \dl{p_c}\cF^\prime
\,,\quad \ldots\,.
\end{equation} 
where $\cF^\prime$ is linearly related to the undifferentiated Fronsdal tensor. 

To make contact with the literature
mention that the extra (with respect to the on-shell version) term in $Q\cC_a$ is related to the certain  $\sigma_-$-cohomology~\cite{Lopatin:1988hz,Bekaert:2005vh} class, namely so-called ``Einstein cohomology'', of the operator $\sigma_-=\xi^a\dl{y^a}$ restricted to act on the space of polynomials in $y^a,p^a$ satisfying $\sd \chi=T\chi=\Box\chi=S\chi=0$. From the present perspective $\sigma_-$ can be identified with $Q$ restricted to the submanifold $\cF=\hat R=0$, known as the gauge module in the unfolded approach.\footnote{More precisely, if $q$ denotes restriction of $Q$ to the subspace then $\sigma_-$ is an associated ``first-quantized BRST operator'' in the sense that $\sigma_-\cC=q\cC$ where $\sigma_-$ act on $y,a,\xi$ while $q$ on coordinates  $\cC^{b_1\ldots b_{s-1}}_{a\ldots}$.}

\subsection{Frame-like Lagrangian from parent formulation}

According to the general prescription of Section~\bref{sec:parent}
to construct a parent formulation in addition to $\tilde\gamma=-d_H+\gamma$ defined on $\manN$
we need a Lagrange potential $\hat L$ which is a representative of the
Lagrangian in the cohomology of $\tilde\gamma$. To compute $\hat L$ and to construct
the parent formulation it is convenient to first eliminate some trivial pairs for
$\gamma$. More precisely, all the components in $C$ and $\phi$ except for those
parametrizing the subspace singled out by $SC=0$ and $T\phi=0$ form contractible
pairs for $\gamma$. Indeed $\gamma T\phi=S C$. This is a jet-space counterpart
of the well-known traceless gauge for the Fronsdal system~\cite{Skvortsov:2007kz} (see also \cite{Alvarez:2006uu,Blas:2007pp}.  Note that the expression for both $d_H$ and $\gamma$
are unchanged as both vector fields reduce to the surface $\manN^\prime \subset \manN$
singled out by $SC=T\phi=0$.

After elimination the Lagrangian takes a simple form~\cite{Skvortsov:2007kz}
\begin{equation}
L=\half\inner{\phi_a}{\phi_a}-\half\inner{S\phi}{S\phi}|_{y=0}\,.
\end{equation} 
We search for a $\tilde\gamma$-invariant completion of $L$ in the form
\begin{equation}
 \hat L=\Vol L+\Vol_a J^a+\half\Vol_{ab} J^{ab}\,.
\end{equation} 
where 
\begin{equation}
 \Vol_{a_1\ldots a_k}=\frac{1}{(n-k)!}\,\xi^{b_1}\ldots \xi^{b_{n-k}}\,\epsilon_{b_1\ldots b_{n-k}a_{a_1}\ldots a_k}\,.
\end{equation} 
Condition $\tilde\gamma\hat L=0$ gives $\gamma L=-\d_a J^a$ and $\gamma J^a=\d_b J^{ba}$. Direct computation gives
a possible solution for $J^a, J^{ab}$:
\begin{equation}
J^a=\inner{\phi}{p^a\Box C}|_{y=0}-\inner{\phi}{\d^a\sd C}|_{y=0}\,,
\end{equation} 
and
\begin{equation}
J^{ba}=\half\left[\inner{p^bC}{p^a\Box C-\d^a \sd C}|_{y=0}-\inner{\sd C}{p^b\d^a C}|_{y=0}-(a \leftrightarrow b)\right]\,,
\end{equation} 
so that we have explicitly constructed all the ingredients for the parent Lagrangian formulation for Fronsdal system.
It is then given by an AKSZ sigma model with the target space $T^*[n-1]\manN^\prime$.

Note that the structure of the $n$-form component $L$ of Lagrange potential $\hat L$ coincides with the Fronsdal Lagrangian in the traceless gauge~\cite{Skvortsov:2007kz} (see also \cite{Alvarez:2006uu,Blas:2007pp}). However, in our setting imposing this gauge doesn't produce differential constraints on the true gauge parameters of the parent formulation because this only restricts the target space so that the constraints are algebraic. This is in contrast to the usual treatment~\cite{Alvarez:2006uu,Blas:2007pp,Skvortsov:2007kz} of the traceless gauge. 
Moreover,
the parent formulation can be build starting from a theory where gauge parameters are subject to differential constraints. In this case, however, the equivalence is not guaranteed and has to be studied separately. If it
happens to be equivalent the parent formulation gives a systematic way to replace the constrained formulation with the
unconstrained one.

We now reduce the system by eliminating contractible pairs for $\tilde\gamma$. According to the discussion in the previous section all the coordinates on $\manN$ (and hence on $\manN^\prime$) form contractible pairs for $\tilde\gamma$ except for $z^a,\xi^a,\cC,\hat R,\cF$ and differential $\tilde\gamma$ induces a reduced differential $Q$ in the space $\manN_{\red}$ of these variables. In what follows we restrict to the case $s\geq 2$
as the reduction for $s=0,1$ is different and was discussed in details in~\cite{Grigoriev:2010ic}.

Upon the reduction $\hat L$ reduces to a representative $\hat L_{red}$ such that
$Q\hat L_{red}=0$. The Lagrangian potential $\hat L$ and hence $\hat L_{red}$ is not unique
as it is defined modulo $\tilde\gamma$-exact or $Q$-exact terms respectively. To understand the
relation with the frame-like formulation a useful choice of $\hat L_{red}$ is to have
it $\hat R, F$-independent. To find such $\hat L_{red}$ we use another representative for $\hat L$. Namely we observe that
the term containing $\inner{\sd C}{\,\cdot\,}$ in $J^{ab}$ can be absorbed by adding $\d_b\inner{\phi}{\,\cdot\,}$ to $J^a$
and hence $-\gamma\inner{\phi}{\,\cdot\,}$ to $J^{ab}$. Adding such term does not affect the reduction of $J^a$
because it only involves 0-th and 1-st derivatives of $\phi$ and hence can not produce either $F$ or $\hat R$ fields
if $s\geq 0$. The first term $\Vol L$ as well as $\Vol_a J^a$ reduce to zero by the same reasoning so that we concentrate on the last term $\half \Vol_{ab} J^{ab}$.

By construction $J^{ab}$ is defined modulo dual total derivative. By adding $\half\d_c T^{bca}$ with 
\begin{equation}
T^{bca}=\inner{p^bC}{(p^c C_a-p^aC_c}+cycle(bca)\,,
\end{equation} 
one finds new representative
\begin{equation}
 {J^\prime}^{ab}=\half\left[\inner{p^aC_d}{p^d C_b-p^b C_d}-(a\leftrightarrow b)\right]+\inner{\sd C}{(p^b\d^a-p^a\d^b)C}|_{y=0}\,.
\end{equation} 
The last term can also be removed by redefining $J^a$ by terms independent on $\hat R,F$.
Finally, putting to zero all the coordinates except for $z^a,\xi^a,F,\hat R,\cC$ 
one finds:
\begin{equation}
 \hat L_{red}=\half \Vol_{ab}\left[\inner{\cC_a}{\cC_b}-\inner{p^a \cC_d}{p^b \cC_d}\right]\,.
\end{equation} 
By construction it is $Q$-invariant. This can be easily checked directly. The reduced parent formulation
is completely determined in terms of $Q$ and $\hat L_{red}$ defined on $\manN_\red$.

We now show that a certain further reduction results in the well-known frame-like Lagrangian. 
We first fix a reparametrization invariance by $z^a=x^a$ and $\xi^a=\theta^a$. Observe then 
the set of fields of the reduced formulation is formed by component fields of $\cC_{(a)},F,\hat R$ and their conjugate antifields. If $\hat L_{red}$ were zero all of the fields except for those originating from $\cC$ and their conjugate 
antifields were generalized auxiliary because the theory would be dynamically empty.
But the Lagrangian potential $\hat L_{red}$ depends on $\cC_{a}$ only  so that all the other fields
along with their antifields are generalized
auxiliary and can be eliminated. Restricting to the classical action and hence putting the antifields  to zero  the action takes the form
 \begin{equation}
 S_R[e,\omega,\Lambda]=\int \inner{\Lambda}{\derham e-\sigma \omega} + \hat L_{red}(\omega)\,,
\end{equation} 
where $e$ and $\omega$ denotes $1$-form fields of vanishing ghost degree originating from target space coordinates  $C$ and $C_a$ respectively. Namely,
\begin{equation}
\begin{aligned}
C(x,\theta,p)&=\overset{0}C(x|p)+\theta^b e_b(x,p)+\half\theta^d\theta^b\overset{2}C_{db}(x|p)+ \ldots\,,\\
C_a(x,\theta,p)&=\overset{0}C{}_a(x|p)+\theta^b \omega_{a|b}(x,p)+\half\theta^d\theta^b\overset{2}C{}_{a|db}(x|p)+\ldots
\end{aligned}
\end{equation} 
In its turn, field $\Lambda$ is an $n-2$-form in the space dual to that where $\overset{0}C$ takes values. It enters the formalism as an antifield conjugate to $\overset{2}C_{ab}$. 

Our next aim is to eliminate $\omega_{a|b}(x,p)$ entering $C_a(x,\theta,p)$ as $\theta^a\omega_{a|b}y^b$. To this end we first note that in the reduced parent formulation $\omega$ is subject to the algebraic gauge symmetry $\delta \omega = \derham (\overset{0}C{}_ay^a)-\sigma( \half\overset{0}C_{ab}y^ay^b)$,
where the ghost fields are to be understood as gauge parameters. Using the symmetry one can assume that $\omega$  doesn't have an irreducible  component whose tensor structure is identical to that of $\overset{0}C_{ab}$
(recall that $\overset{0}C_{ab}$ is totally traceless and $p^a\overset{0}C_{ab}=0$). After gauge fixing the space of such $\omega$ is isomorphic to the space of $\Lambda$.

It is convenient to introduce a new inner product for elements  of tensor structure as $\omega$
(which we write as generating functions of $y,p,\theta)$) as follows
\begin{equation}
 \Vol\inner{F_ay^a}{G_by^b}^\prime=\Vol_{cab} \inner{\bar p^c F_a}{\bar p^b \theta^d G_d}\,.
\end{equation} 
which is nothing but the inner product determining the quadratic part of the
frame-like action of~\cite{Vasiliev:1980as}. Note that the inner product is
nondegenerate. In terms of $\inner{}{}^\prime$ the expression for
$L_{red}(\omega)$ takes the form $
L_{red}(\omega)=\half\Vol\inner{\omega}{\omega}^\prime$.  It is also convenient
to parametrize fields in $\Lambda$ in terms of $\hat \omega$ such that
$\hat\omega$ has the same tensor structure as $\omega$ and
$\inner{\Lambda}{\omega}=\Vol\inner{\hat\omega}{\omega}^\prime$. 

In these terms $S_R$ takes the  form
\begin{equation}
 S_R[e,\omega,\hat\omega]=\int  d^n x (\inner{\hat\omega}{y^a\dl{x^a}e -\omega}^{\prime}
+\half\inner{\omega}{\omega}^\prime)\,.
\end{equation} 
Varying with respect to $\omega$ gives $\omega=\hat\omega$. Then eliminating
$\omega$ gives the familiar frame-like action 
\begin{equation}
 S_{frame}[e,\hat\omega]=
\int  d^n x \,\inner{\hat\omega}{y^a\dl{x^a}e -\half \hat\omega}^{\prime}
=
\int d^n\theta d^n x \, \Vol_{cab}
\inner{\bar p^c \hat\omega_a}{\bar p^b(\derham e-\half\sigma \hat\omega)}
\end{equation} 
of~\cite{Vasiliev:1980as}. It is natural to expect that the present approach can also help in understanding
the structure underlying more general frame-like Lagrangians~\cite{Zinoviev:2003dd,Skvortsov:2008sh,Zinoviev:2009gh}.

Note that the above procedure can be used to explicitly relate cubic vertexes in the metric-like and the frame-like   formulations. Indeed, using in the above procedure the deformed Lagrangian as a staring point one should end up with the respective deformation of the frame-like Lagrangian. This can be e.g. used to explicitly relate the metric-like vertexes (see~\cite{Metsaev:2005ar,Manvelyan:2010wp,Sagnotti:2010at,Fotopoulos2010a,Manvelyan:2010je} for a complete description and references to earlier contributions) and the frame-like ones of~\cite{Vasilev:2011xf} and Refs. therein.

\subsection{Off-shell constraints and gauge symmetries at the nonlinear level}
\label{sec:off} 

Working at the level of equations of motion let us consider the off-shell version
of the parent system  \eqref{eom-P}-\eqref{gs-P}
where in contrast to considerations in Section~\bref{sec:F-eom}
fields $A,F$ and gauge parameters are not subject to any constraint. In particular
they are traceful and are not of definite homogeneity in $p_a$ so that the system
describes fields of all integer spins. In addition we work with the non-parametrized version where
$z^a=x^a$ and $\xi^a=\theta^a$.

It was observed in~\cite{Vasiliev:2005zu} that this system is a linearization of
\begin{equation}
\label{nlin}
 dA+\half\qcommut{A}{A}=0\,, \qquad dF+\qcommut{A}{F}=0\,,
\end{equation}    
around a particular solution
\begin{equation}
A_0=\theta^b p_b, \qquad  F_0=\half\eta^{ab}p_ap_b\,.
\end{equation}
This can be easily checked using 
$\qcommut{A_0}{\cdot}=-\sigma, \quad
\qcommut{\cdot}{F_0}=\sd$. 
Here $\qcommut{\cdot}{\cdot}$ denotes the Weyl $*$-commutator determined by
$\qcommut{y^a}{p_b}=\delta^a_b$.  The above nonlinear system can be related to
the master equation for the quantized scalar particle propagating in the higher
spin filed background~\cite{Grigoriev:2006tt} (see also~\cite{Segal:2002gd} for the closely related interpretation in terms of the conformal fields).

The AdS-space version of the system~\eqref{nlin} which also takes into account
the double-tracelessness condition is also available~\cite{Grigoriev:2006tt}.
However, the familiar $sp(2)$-symmetry of the full nonlinear system from~\cite{Vasiliev:2003ev}
is not manifest in that proposal. We now give a natural generalization of~\eqref{nlin} to the AdS case which has both $sp(2)$ and AdS invariance manifest.

We recall that the $AdS_n$ space $\manX_0$ with coordinates $x^\mu$ can be described in terms of the
$o(n-1,2)$-vector bundle with the fiber $\fR^{n+1}$ and equipped with the flat $o(n-1,2)$ connection $\omega$ and a fixed section $V$ satisfying $V^AV_A=-1$. This bundle can be seen as a pullback of the
tangent bundle over the ambient space $\fR^{n+1}$ by the embedding of $\manX_0$ as a hyperboloid in $\fR^{n+1}$.
The connection originates from the standard metric connection on the ambient space and $V$ from the tautological section of the ambient tangent bundle. For more details see e.g.~\cite{Bekaert:2005vh,Barnich:2006pc}.

The system we are searching for is directly constructed as an AKSZ sigma model.
Let $Y^A$ be coordinates on $\fR^{n+1}$ and $P_A$ their dual. Let $\algg$ be $sp(2)$-algebra
and $e_1,e_2,e_3$ its standard basis i.e. $[e_2,e_1]=2e_1$, $[e_2,e_3]=-2e_3$
and $[e_1,e_3]=e_2$. Introduce coordinates $\nu^i$ on $\Pi\algg$ and take
$\gh{\nu^i}=1$. These are naturally interpreted as ghosts for the BRST realization
of the Lie algebra complex so that the differential is $q=-\half \nu^i\nu^jU_{ij}^k
\dl{\nu^k}$, where $U_{ij}^k$ are structure constants of $\algg$.

The target space supermanifold $\manM$ is introduced as follows. 
Consider the algebra $\cA$ of polynomials in variables $P_A$ and $\nu^i$ with coefficients in formal series in
$Y^A$ and take as $\manM$ vector space $\cA$ with reversed parity and
shifted ghost degree. More precisely, coordinates on $\manM$ are coefficients
of the generating function
\begin{equation}
 \Psi(c,Y,P)=C(Y,P)+\nu^i F_i(Y,P)+\nu^i\nu^j G_{ij}(Y,P)+\nu^i\nu^j\nu^k G_{ijk}(Y,P)\,.
\end{equation} 
The ghost degree and parity of the coordinates are determined by $\gh{\Psi}=\p{\Psi}=1$. In particular,
$\gh{C}=1$, $\gh{F_i}=0$, $\gh{G_{ij}}=-1$, $\gh{G_{ijk}}=-2$ and the Grassmann parity is the ghost degree modulo $2$.

We need two structures on $\cA$: the Weyl star-product determined by $\qcommut{Y^A}{P_B}=\delta^A_B$
and the Lie algebra differential $q=-\half \nu^i\nu^jU_{ij}^k \dl{\nu^k}$.
These structures induce on $\manM$ an odd vector field $Q$ determined by~\footnote{It is not difficult to see that $\manM$ is a product of two $Q$-manifolds. One is $\Pi \algg$ equipped with $q$ and another one is $\Pi(\text{Weyl commutator algebra})$
 equipped with the respective Lie algebra differential. $Q$ is simply a product $Q$-structure. 
}
\begin{equation}
 Q\Psi=q\Psi+\half\qcommut{\Psi}{\Psi}\,.
\end{equation} 
It is nilpotent $Q^2=0$ and of ghost degree $1$. The nilpotency is a consequence of $q^2=0$,
Jacobi identity for $\qcommut{\cdot}{\cdot}$, and that $q$
differentiates $\qcommut{\cdot}{\cdot}$. In coordinate terms one has
\begin{equation}
\begin{gathered}
QF_{i}=\qcommut{F_{i}}{C}\,, \qquad\qquad 
QC=\half\qcommut{C}{C}\,, \\\qquad
QG_{ij}=\half\qcommut{F_{i}}{F_{j}}- \half U_{ij}^k F_k+\qcommut{G_{ij}}{C}\,,\qquad \ldots\,.
\end{gathered}
\end{equation} 
Here we do not write the analogous relations for the remaining coordinates because they are not needed 
for the equations of motion and gauge symmetries
and only serve the BRST formulation.

Consider AKSZ sigma model with the target space $\manM$ and the source
$\manX=\Pi T \manX_0$ with coordinates $x^\mu,\theta^\mu$ (recall that $\manX_0$ is an AdS space).
Equations of motion take the form
\begin{equation}
\label{eom}
 \derham A+\half\qcommut{A}{A}=0\,, \quad \derham F_i+\qcommut{A}{F_i}=0\,, \qquad
\qcommut{F_{i}}{F_{j}} - U_{ij}^k F_k=0\,,
\end{equation} 
where by some abuse of notations we introduced component fields of ghost degree $0,1$
according to $C(x,\theta|Y,P)=C(x|Y,P)+\theta^\mu A_\mu(x|Y,P)+\ldots$ and
$F_i(x,\theta|Y,P)=F_i(x|Y,P)+\theta^\mu \ldots$.
The gauge transformations are
\begin{equation}
\label{gs}
 \delta_\lambda F_i=\qcommut{F_i}{\lambda}\,,\qquad \delta_\lambda A=\derham\lambda+\qcommut{A}{\lambda}\,,
\end{equation} 
where $\lambda(x|Y,P)$ is a gauge parameter replacing the ghost field $C(x|Y,P)$.

As a next step we analyze the linearization of~\eqref{eom} and \eqref{gs}
around the following background solution: 
\begin{equation}
\Psi_0=\theta^\mu A^0_\mu+\nu^i F^0_i\,, \\
\end{equation} 
where
\begin{equation}
 \begin{gathered}
A^0_\mu=\theta^\mu \omega_{\mu A}^B(x)(Y^A+V^A)P_B\,,\\
F^0_1=P^A P_A\,, \quad F^0_2=(Y^A+V^A)P_A\,, \quad F^0_3=-(Y^A+V^A)(Y_A+V_A)\,.
\end{gathered}
\end{equation} 
Here $\omega_{\mu A}^B(x)$ are coefficients of the flat AdS connection and $V^A$ are components of the compensator, and
indexes are contracted using the $o(d-1,2)$-invariant metric. In addition, we assume that the local frame is such that
$V^A=\const$.

It follows that the linearized system is precisely the off-shell version of the  parent formulation
from~\cite{Barnich:2006pc} for totally symmetric AdS gauge fields. To see this note that the linearized equations and gauge symmetries can be encoded in the following BRST operator\footnote{
Note that this BRST operator can be directly obtained by
linearizing the AKSZ sigma model BRST differential around the particular solution using the 
technique of~\cite{\Goff} (see Section~3.2. of that Ref.).}

\begin{equation}
\Omega=\derham+\qcommut{A^0}{\cdot}+\nu^i\qcommut{F^0_i}{\cdot}+q\,,
\end{equation} 
acting on the space of states $\Psi(x,\theta|Y,P,c)$. One then identifies 
\begin{equation}
 \derham+\qcommut{A^0}{\cdot}=\derham+
\theta^\mu \omega_{\mu A}^B\left(P_B\dl{P_A}-(Y^A+V^A)\dl{Y^B}\right)
\end{equation} 
as the covariant derivative in the so-called twisted realization (see~\cite{Barnich:2006pc,Alkalaev:2009vm}
for more details). Furthermore,
\begin{multline}
 \nu^i\qcommut{F^0_i}{\cdot}+q=\\=-\nu^1 P^A\dl{Y^A}+\nu^2(P_A\dl{P_A}-(Y^A+V^A)\dl{Y^A})-\nu^3 (Y^A+V^A)\dl{P^A}+q
\end{multline}
gives the fiber part which is the BRST operator of $sp(2)$ represented on $Y,P$ variables. The only difference
with~\cite{Barnich:2006pc} is that fields are traceful and $\nu^1$ is represented in the coordinate rather than 
the momenta representation. Assuming $\Psi(x,\theta|Y,P,c)$ totally traceless results in the on-shell system of~\cite{Barnich:2006pc} which is equivalent to the Fronsdal equations. Note that requiring $\Psi$ traceless imposes constraints on both fields and gauge parameters (for low spin components the gauge parameter constraints are identically satisfied). The same applies to the off shell-systems from~\cite{Vasiliev:2005zu,Grigoriev:2006tt}.
This is in contrast to e.g. off-shell system~\cite{Segal:2002gd} for conformal fields where the on-shell version is obtained by imposing constraints on fields only.

\noindent
The following comments are in order: 

-- The construction can be easily generalized by replacing $\algg$ with a generic Lie algebra. One can check that the system remains consistent. In this way we actually find a family of consistent systems.
For instance taking $\algg=\fR^1$ and choosing $\manX_0$ to be the ambient space itself (in this case it is natural to identify it as $n+1$-dimensional Minkowski space) reproduces flat space nonlinear system~\eqref{nlin} from~\cite{Vasiliev:2005zu}.

-- Among the target space coordinates one finds $G_{ij},G_{ijk}$ which carry negative ghost degree so that the system is not just an FDA so that using the parent formulation formalism is essential in this case. Indeed one and the same homological vector field $Q$ encodes both the FDA relations and the constraints in the consistent way.

-- Just like~\eqref{nlin} the general nonlinear system can also be interpreted as a specific BFV master equation for a quantized particle propagating in higher spin background in the spirit of~\cite{Grigoriev:2006tt}. 

-- Using the technique from~\cite{Alkalaev:2009vm} the system can be reformulated in the purely ambient-space terms where fields do not depend on $Y$-variables.  This can be obtained by replacing $Y^A+V^A\to X^A$ and dropping $x^\mu,\theta^\mu$. The only remaining fields are $F(X|P)$ entering $\Psi$ as $\Psi=\eta^i F_i$ and the equations of motion are simply the $sp(2)$ relations $\qcommut{F_i}{F_j}=U_{ij}^k F_k$ where now $\qcommut{X^A}{P_B}=\delta^A_B$.
This reformulation, however, is not a strictly local operation. 

-- The system admits a natural truncation to low spin sector. Indeed, consider the following degree:
$\deg Y^A=0$, $\deg P_A=1$. In the sector of ghost variables take $\deg \nu^1=-1$, $\deg \nu^2=0$, $\deg \nu^3=1$
(i.e. the degree is proportional to the weight of the respective generators if $\algg$ is seen as the adjoint $sp(2)$-module). It is easy to see that $\deg{\qcommut{}{}}\leq -1$ and $\deg{q}=0$ so that elements of 
$\cA$ of degree less or equal than $1$ form a subspace $\cA_0$ invariant under both $q$ and $\qcommut{}{}$.
The system can be then consistently trancated to that associated with $\cA_0$ by simply requiring $\deg{\Psi}\leq 1$.
By examining the linearized version this can be seen to describe spins $0,1,2$. One can also describe spin $2$ alone by taking $\deg{\Psi}=1$ and replacing $\qcommut{}{}$ with the respective Poisson bracket.

-- A natural question is to find a nonlinear version of trace constraints in order to construct a
parent formulation of the Vasiliev nonlinear system~\cite{Vasiliev:2003ev}.

\section*{Acknowledgments}
\label{sec:acknowledgements}
Useful discussions with K.~Alkalaev, G.~Barnich, I.~Batalin, D.~Francia, S.~Lyakhovich, E.~Skvortsov, A.~Smirnov, M.~Vasiliev, and A.~Verbovetsky are gratefully acknowledged. This work was supported by RFBR grant 11-01-00830.

\addtolength{\baselineskip}{-5.8pt}
\providecommand{\href}[2]{#2}\begingroup\raggedright\endgroup

\end{document}